\documentclass[11pt,a4paper]{article}
\pdfoutput=1
\usepackage{jcappub}
\usepackage{rotating}
\usepackage{array}
\usepackage{amsmath}
\usepackage{slashed,hyperref}
\usepackage{caption}
\usepackage[normalem]{ulem}
\usepackage{afterpage}
\usepackage{placeins}

\newcommand{\td}{\ensuremath{\text{d}}}

\def\bm#1{\boldsymbol{#1}}
\newcommand\RR[3]{\mathrm{RR}_\mathrm{#1}(\text{#2}/\text{#3})}

\title{Studying generalised dark matter interactions with extended halo-independent methods}

\author[a]{Felix Kahlhoefer}
\author[b]{and Sebastian Wild}

\affiliation[a]{DESY, Notkestra\ss e 85, D-22607 Hamburg, Germany}
\affiliation[b]{Physik-Department T30d, Technische Universit\"at M\"unchen,\\James-Franck-Stra\ss e 1, D-85748 Garching, Germany}

\emailAdd{felix.kahlhoefer@desy.de}
\emailAdd{sebastian.wild@ph.tum.de}

\abstract{The interpretation of dark matter direct detection experiments is complicated by the fact that neither the astrophysical distribution of dark matter nor the properties of its particle physics interactions with nuclei are known in detail. To address both of these issues in a very general way we develop a new framework that combines the full formalism of non-relativistic effective interactions with state-of-the-art halo-independent methods. This approach makes it possible to analyse direct detection experiments for arbitrary dark matter interactions and quantify the goodness-of-fit independent of astrophysical uncertainties. We employ this method in order to demonstrate that the degeneracy between astrophysical uncertainties and particle physics unknowns is not complete. Certain models can be distinguished in a halo-independent way using a single ton-scale experiment based on liquid xenon, while other models are indistinguishable with a single experiment but can be separated using combined information from several target elements.}


\arxivnumber{DESY-16-132, TUM-HEP 1053/16}

\begin{document}

\maketitle

\flushbottom

\section{Introduction}

Over the past decade dark matter (DM) direct detection experiments have improved their sensitivity by an order of magnitude every two years and this trend is expected to continue for the near future~\cite{Cushman:2013zza,Baudis:2014naa,Akerib:2015cja,Aprile:2015uzo,Aalbers:2016jon,PICO250,Shields:2015wka,Calkins:2016pnm,Arnaud:2016tpa}. While there is at present no clear evidence for the interactions of DM particles in nuclear recoil detectors~\cite{Akerib:2015rjg,Aprile:2012nq,Amole:2016pye,Agnese:2014aze,Angloher:2015ewa}, it is perfectly conceivable (and in fact predicted by many models for DM) that hundreds of events will be observed by 2020. Once a signal is seen in one or several direct detection experiments, the challenge will be to identify those models of DM that allow for a good fit of the experimental data and to determine the preferred values of the underlying parameters, such as the mass of the DM particle.

Answering these questions is complicated by the fact that event rates in direct detection experiments depend in complicated ways on the velocity distribution of DM particles in the Galactic halo, which is subject to large uncertainties~\cite{Kuhlen:2009vh,Lisanti:2010qx,McCabe:2010zh,Mao:2012hf}. Many studies have investigated the impact of these uncertainties on our ability to employ the results of future direct detection experiments in order to infer the mass of the DM particle or to discriminate between different DM models (e.g.\ by distinguishing between spin-independent and spin-dependent interactions or by determining separately the DM-proton and the DM-neutron coupling)~\cite{Strigari:2009zb,Peter:2009ak,Green:2010gw,Peter:2011eu,Pato:2012fw,Kavanagh:2012nr,Frandsen:2013cna,Kavanagh:2013wba,Peter:2013aha,Strigari:2013iaa,Kavanagh:2014rya,Feldstein:2014gza,Feldstein:2014ufa}. In particular, a range of so-called halo-independent methods have been developed to reduce or even eliminate completely the impact of astrophysical uncertainties on the DM properties that can be inferred from direct detection experiments~\cite{Drees:2008bv,Fox:2010bu,Fox:2010bz,McCabe:2011sr,Frandsen:2011gi,Gondolo:2012rs,HerreroGarcia:2012fu,DelNobile:2013cva,Fox:2014kua,Feldstein:2014gza,Feldstein:2014ufa,Bozorgnia:2014gsa,Cherry:2014wia,Anderson:2015xaa,Gelmini:2015voa,Ferrer:2015bta,Scopel:2015baa,DelNobile:2015rmp,Gelmini:2016pei}.

At the same time, however, it has become clear that the interactions of DM particles with nuclei can be significantly more complicated than suggested by the simple division into spin-independent and spin-dependent interactions. In general the scattering cross section can also depend on the momentum transfer and the relative velocity between the DM particle and the nucleus. A full classification of all possible DM interactions in the non-relativistic limit requires no less than 28 different scattering operators as well as a large number of nuclear response functions~\cite{Fitzpatrick:2012ix,Anand:2013yka}, which can significantly affect the expected signals in direct detection experiments~\cite{Fitzpatrick:2012ib,Catena:2014uqa,Gresham:2014vja,Catena:2014hla,Catena:2014epa,Gluscevic:2014vga,Catena:2015uua,Gluscevic:2015sqa,Scopel:2015baa,Dent:2015zpa,Catena:2016hoj}. The possible importance of the additional DM-nucleon scattering operators makes the issue of astrophysical uncertainties even more pressing, because a non-standard velocity distribution can potentially mimic non-standard DM interactions. 

Fortunately, the degeneracy between astrophysical uncertainties and particle physics unknowns is not complete. For example, standard spin-independent interactions induce a differential event rate  that decreases monotonically with increasing recoil energy for \emph{any} DM velocity distribution~\cite{Fox:2010bu}. The observation of a non-monotonic differential event rate could hence not be attributed to astrophysical uncertainties and would instead point strongly towards non-standard interactions. It should therefore always be possible to obtain at least some basic information on the nature of DM interactions even when accounting for astrophysical uncertainties.

In the present paper we develop a framework to combine the full formalism of non-relativistic effective interactions with state-of-the-art halo-independent methods. Our approach is based on the idea of decomposing the velocity distribution into a finite sum of streams with velocity $v_j$ and then varying the normalisation of each stream~\cite{Feldstein:2014gza,Feldstein:2014ufa}. We will show that, even for non-standard interactions, it is always possible to calculate a matrix $D_{ij}$ such that the number of events $N_i$ in the $i$th bin of a given experiment can be calculated via simple matrix multiplication $N_i = D_{ij} \, g_j$ with the $g_j$ being determined by the normalizations of the streams. This simple relation allows to determine the velocity distribution that best describes a given set of data for assumed particle physics properties.

We can repeat this procedure for different particle physics assumptions in order to study whether changes in the velocity distribution can compensate for changes in the particle properties of DM and thus reduce our ability to determine these properties unambiguously. The aim is to quantify what information can be inferred on the coupling structure of DM in a halo-independent way. For example, if no good fit to a given set of data can be found for any DM velocity distribution, the corresponding particle physics assumptions can be disfavoured without the need to make assumptions on the astrophysical distribution of DM. This approach makes it possible to analyse direct detection experiments for arbitrary DM interactions, in particular DM interactions with non-standard momentum and velocity dependence, independent of astrophysical uncertainties.

To illustrate the general formalism we study a representative set of DM-nucleon interactions. These consist of the standard spin-independent and spin-dependent scattering scenarios, interactions induced by an anapole or a magnetic dipole moment of DM, as well as a dipole interaction involving a new heavy mediator. This set of interactions, though certainly not exhaustive, covers all the central aspects relevant for a halo-independent investigation of non-standard interactions between DM and nucleons. We discuss the scattering rates induced by these models in the context of three future direct detection experiments based respectively on xenon, germanium and iodine targets. We first study whether a single (xenon-based) experiment can distinguish the different models of DM in a halo-independent way and then focus on the question whether the complementarity of several different target materials can improve the distinction.

A similar analysis of the possibility to distinguish different DM models using future direct detection experiments has been performed in~\cite{Gluscevic:2015sqa}. While considering a larger set of DM interactions and more different experiments, the analysis makes much more specific assumptions on the velocity distribution of DM. Our results extend the findings from~\cite{Gluscevic:2015sqa} to arbitrary DM velocity distributions (and furthermore take into account effects from finite energy resolution).

This paper is structured as follows. In section~\ref{sec:haloindependent} we review the basic formalism for direct detection, including the central ideas underlying the halo-independent methods for DM-nucleon interactions with standard velocity dependence. We then explain how to generalise such a framework to more complicated DM interactions and derive the central formulas used to calculate the matrix $D_{ij}$. In section~\ref{sec:generalised}, after defining the set of interaction scenarios discussed in this work, we provide a qualitative illustration of the halo-independent interpretation of direct detection data in the context of these models. We then introduce methods for quantifying the goodness of halo-independent fits to experimental data. In sections~\ref{sec:1exp} and \ref{sec:2exp} we apply the method to various combinations of direct detection experiments, and discuss in particular which of the above-mentioned interaction scenarios can be distinguished in a halo-independent way. Section~\ref{sec:conclusions} provides a summary of our findings. Additional details on the parametrization of non-relativistic effective interactions and the calculation of event rates are given in the appendices.

\section{Extended halo-independent methods}
\label{sec:haloindependent}

This section describes how the integral of the DM velocity distribution can be used to study astrophysical uncertainties in direct detection experiments. After briefly recalling the formalism of writing the event rates in terms of velocity integrals, we review the approach for analysing direct detection data in a halo-independent way for the case where the differential cross section has the standard velocity dependence $\mathrm{d}\sigma/\mathrm{d}E_\mathrm{R} \propto v^{-2}$. Crucially for the remainder of this work, we then generalise the method to more complicated cross sections. Additional technical details related to this section can be found in the appendices.

\subsection{Calculating event rates from velocity integrals}

The differential event rate with respect to recoil energy in a direct detection experiment is given by
\begin{equation}
\frac{\text{d}R}{\text{d}E_\text{R}} = \frac{\rho}{m_T \, m_\chi}  \int_{v_\text{min}}^\infty v f(\boldsymbol{v} + \boldsymbol{v}_\text{E}(t)) \frac{\text{d} \sigma}{\mathrm{d} E_{\text{R}}}  \text{d}^3 v\; ,
\label{eq:dRdE}
\end{equation}
where $\rho$ is the local DM density, $m_\chi$ and $m_T$ are the DM and target nucleus mass, $f(\mathbf{v})$ is the local DM velocity distribution evaluated in the Galactic rest frame, $\boldsymbol{v}_\text{E}(t)$ is the velocity of the Earth relative to the Galactic rest frame and $v=|\boldsymbol{v}|$. For elastic scattering the \emph{minimum velocity} required for a DM particle to transfer the energy $E_\text{R}$ to a given nucleus $T$ is
\begin{equation}\label{eq:vmin}
v_\text{min}(E_\text{R}) = \sqrt{\frac{m_T E_\text{R}}{2 \, \mu_{T\chi}^2}} \;,
\end{equation}
where $\mu_{T\chi} = m_T \, m_\chi / (m_T + m_\chi)$ is the reduced mass of the DM-nucleus system.

The differential cross section $\text{d} \sigma / \mathrm{d} E_{\text{R}}$ encodes the details of the interactions between DM particles and nuclei. As discussed in detail in Appendix~\ref{ap:non-relativistic}, for a very general set of effective operators, the differential cross section can always be decomposed into a contribution proportional to $v^{-2}$ and a velocity-independent contribution:
\begin{equation}
\frac{\text{d} \sigma}{\mathrm{d} E_{\text{R}}} = \frac{\text{d} \sigma_1}{\mathrm{d} E_{\text{R}}} \frac{1}{v^2} + \frac{\text{d} \sigma_2}{\mathrm{d} E_{\text{R}}} \; .
\end{equation}
Defining the velocity integrals\footnote{Here we make two approximations. First, we neglect the small time dependence of $\mathbf{v}_\text{E}$ and second we assume that $\text{d} \sigma/\mathrm{d} E_{\text{R}}$ only depends on $v^2$ and not on the direction of $\mathbf{v}$ (e.g.\ because the target is unpolarised and the direction of recoil tracks cannot be distinguished).}
\begin{equation}
 g(v_\text{min}) = \int_{v_\text{min}}^\infty \frac{1}{v} \, f(\boldsymbol{v} + \boldsymbol{v}_\text{E}(t)) \, \text{d}^3 v \; , \quad  h(v_\text{min}) = \int_{v_\text{min}}^\infty v \, f(\boldsymbol{v} + \boldsymbol{v}_\text{E}(t))\, \text{d}^3 v
\end{equation}
we can then simply write the differential event rate as
\begin{equation}
\frac{\text{d}R}{\text{d}E_\text{R}} = \frac{\rho}{m_T \, m_\chi} \left(\frac{\text{d} \sigma_1}{\mathrm{d} E_{\text{R}}} g(v_\text{min}) + \frac{\text{d} \sigma_2}{\mathrm{d} E_{\text{R}}} h(v_\text{min})\right) \; .
\label{eq:dRdEsimple}
\end{equation}

If the second term of this equation is absent, all information on the DM velocity distribution is encoded in the velocity integral $g(v_\text{min})$. As a result, this quantity is particularly convenient for studying the impact of astrophysical uncertainties on observables in direct detection experiments, such as the number of events $R_i$ in a bin of the form $\left[E_i, E_{i+1}\right]$. In order to study how $R_i$ depends on the velocity integral, it was proposed in~\cite{Feldstein:2014gza} to parametrize $g(v_\text{min})$ as a piecewise constant function with $N_\mathrm{s}$ steps.\footnote{Note that this is equivalent to writing the speed distribution $f(v)$ as a sum of $\delta$-functions. In other words, we decompose the velocity distribution into a sum of streams with fixed velocities and negligible velocity dispersion.} By choosing $N_\mathrm{s}$ large enough, this allows for an arbitrarily good approximation to any possible velocity integral.  Concretely, as suggested in~\cite{Feldstein:2014gza}, we divide the region of $v_\text{min}$-space probed by the experiments under consideration into $N_\mathrm{s}$ intervals of the form $\left[v_j,\,v_{j+1}\right]$, where the number of steps $N_\mathrm{s}$ can be as large as 50. In each interval, the velocity integral is taken to be constant:
\begin{align}
g(v_\text{min}) = g_j \quad \text{for } \, v_\text{min} \in \left[v_j, v_{j+1} \right] \,.
\label{eq:def_gj}
\end{align}
The fact that the velocity integral is both positive and monotonically decreasing leads to the requirement $0 \leq g_j \leq g_{j-1}$ for all $j$. It was then shown in~\cite{Feldstein:2014gza} that it is possible to encode all experimental details and particle physics properties of DM into a matrix $G_{ij}$ such that $R_i = \sum_j G_{ij} g_j$. Given the very simple relation between the parameters $g_j$ and the observables $R_i$ it is then straight-forward to find the velocity integral that leads to the best agreement with data, effectively profiling out all halo uncertainties.

\subsection{Halo-independent methods for generalised dark matter interactions}

At first sight, the situation becomes more complicated if both $g(v_\text{min})$ and $h(v_\text{min})$ contribute to the differential event rate. The crucial observation is however that $g(v_\text{min})$ and $h(v_\text{min})$ are not independent. In fact, we can make use of the fact that $g'(v) = -f(v)/v$, where $f(v) = \int v^2 \, f(\mathbf{v}+\mathbf{v}_\text{E}) \, \mathrm{d}\Omega_v$, to write~\cite{DelNobile:2013cva,DelNobile:2015rmp}
\begin{align}
 h(v) & = - \int_{v_\text{min}}^\infty v^2 \, g'(v) \mathrm{d}v \nonumber \\
& = \left[-g(v) \, v^2\right]_{v_\text{min}}^\infty + \int_{v_\text{min}}^\infty 2\,v\,g(v) \mathrm{d}v \; .
\label{eq:hfromg}
\end{align}
In other words, it is possible to calculate $h(v_\text{min})$ directly from the usual velocity integral $g(v_\text{min})$.

This observation is key to generalising existing halo-independent methods to models with a more complicated velocity dependence.  To this end, we write the velocity integral as
\begin{equation}
 g(v_\text{min}) = \sum_{j=1}^{N_\mathrm{s}} l_{j} \Theta(v_{j+1} - v_\text{min}) \; ,
\end{equation}
where $\Theta(x)$ is the Heaviside step function and $l_j \equiv g_j - g_{j+1}$, with the $g_j$ defined via equation~(\ref{eq:def_gj}), setting in addition $g_{N_\mathrm{s} + 1} \equiv 0$. We can then use equation~(\ref{eq:hfromg}) to calculate $h(v_\text{min})$:
\begin{equation}
 h(v_\text{min}) = \sum_{j=1}^{N_\mathrm{s}} l_j \, v_{j+1}^2 \, \Theta(v_{j+1} - v_\text{min}) \; ,
\end{equation}
where details of the calculation are provided in appendix~\ref{ap:hcalculation}. Most importantly, we find that $h(v_\text{min})$ is also piecewise constant and monotonically decreasing, i.e.\ we can write $h(v_\text{min}) = h_j$ for $v_\text{min} \in \left[v_j,\,v_{j+1}\right]$ with $0 \leq h_j \leq h_{j-1}$. The parameters $g_j$ and $h_j$ are then related by a simple linear transformation:
\begin{equation}
h_j = \sum_{j'} F_{jj'} g_{j'} \; ,
\end{equation}
with the matrix $F$ given by $F_{jj} = v_{j+1}^2$, $F_{jj'} = 0$ for $j > j'$ and $F_{jj'} = v_{j'+1}^2 - v_{j'}^2$ for $j' > j$ (see also appendix~\ref{ap:hcalculation}).

In order to calculate predicted event rates for a given direct detection experiment we need to know the probability $p(E_\mathrm{D}; E_\mathrm{R})$ that a DM interaction with true (i.e.\ physical) recoil energy $E_\mathrm{R}$ leads to a signal with detected energy $E_\mathrm{D}$. In terms of this function, and using our ansatz for $g(v_\text{min})$, we find
\begin{align}
\frac{\mathrm{d}R}{\mathrm{d}E_\mathrm{D}} & = \int_0^\infty \frac{\text{d}R}{\text{d}E_\text{R}} \, p(E_\mathrm{D}; E_\mathrm{R}) \, \mathrm{d}E_\mathrm{R} \nonumber \\
& = \frac{\rho}{m_T \, m_\chi} \int_0^\infty \left(\frac{\text{d} \sigma_1}{\mathrm{d} E_{\text{R}}} g(v_\text{min}) + \frac{\text{d} \sigma_2}{\mathrm{d} E_{\text{R}}} h(v_\text{min})\right) \, p(E_\mathrm{D}; E_\mathrm{R}) \, \mathrm{d}E_\mathrm{R} \nonumber \\
& = \frac{\rho}{m_T \, m_\chi} \sum_j \left[g_j \int_{E_j}^{E_{j+1}} \frac{\text{d} \sigma_1}{\mathrm{d} E_{\text{R}}} \, p(E_\mathrm{D}; E_\mathrm{R}) \, \mathrm{d}E_\mathrm{R} + h_j \int_{E_j}^{E_{j+1}} \frac{\text{d} \sigma_2}{\mathrm{d} E_{\text{R}}} \, p(E_\mathrm{D}; E_\mathrm{R}) \, \mathrm{d}E_\mathrm{R} \right] \; ,
\end{align}
where $E_j$ is defined implicitly by $v_j = v_\text{min}(E_j)$.

If we are interested in the number of events $R_i$ predicted in a given bin, we simply need to multiply $\mathrm{d}R/\mathrm{d}E_\mathrm{D}$ with the total exposure $\kappa$ and integrate over $E_\mathrm{D}$:
\begin{equation}
R_i = \kappa \int_{E_i}^{E_{i+1}} \frac{\mathrm{d}R}{\mathrm{d}E_\mathrm{D}} \, \mathrm{d}E_\mathrm{D} \; .
\end{equation}
We now make the simplifying assumption that the detection efficiency depends only on the physical recoil energy $E_\mathrm{R}$ (and not on $E_\mathrm{D}$) and that fluctuations can be approximated by a Gaussian distribution with standard deviation given by $\Delta E(E_\mathrm{R})$. In this case we can write
\begin{equation}
p(E_\mathrm{D}; E_\mathrm{R}) = \epsilon(E_\mathrm{R}) \, \frac{1}{\sqrt{2 \pi}  \Delta E(E_\mathrm{R})}\exp\left[-\frac{(E_\mathrm{D} - E_\mathrm{R})^2}{2 \Delta E(E_\mathrm{R})^2} \right] \; .
\end{equation}
The integration over $E_\mathrm{D}$ can then be performed analytically, and one obtains
\begin{equation}
R_i = \sum_j G_{ij} g_j + \sum_k H_{ik} h_k \;,
\label{eq:eventrate}
\end{equation}
where
\begin{align}
G_{ij} & = \frac{\kappa \, \rho}{2 m_T \, m_\chi} \int_{E_j}^{E_{j+1}} \frac{\text{d} \sigma_1}{\mathrm{d} E_{\text{R}}} \, \epsilon(E_\mathrm{R}) \left[\text{erf}\left(\frac{E_{i+1} - E_\text{R}}{\sqrt{2} \Delta E_\text{R}}\right)-\text{erf}\left(\frac{E_i - E_\text{R}}{\sqrt{2} \Delta E_\text{R}}\right)\right] \, \mathrm{d}E_\mathrm{R} \,,
\label{eq:Gij}
\\
H_{ij} & = \frac{\kappa \, \rho}{2 m_T \, m_\chi} \int_{E_j}^{E_{j+1}} \frac{\text{d} \sigma_2}{\mathrm{d} E_{\text{R}}} \, \epsilon(E_\mathrm{R}) \left[\text{erf}\left(\frac{E_{i+1} - E_\text{R}}{\sqrt{2} \Delta E_\text{R}}\right)-\text{erf}\left(\frac{E_i - E_\text{R}}{\sqrt{2} \Delta E_\text{R}}\right)\right] \, \mathrm{d}E_\mathrm{R} \,,
\label{eq:Hij}
\end{align}
with $\text{erf}(x) = (2/\sqrt{\pi}) \int_0^x \exp(-t^2) \, \mathrm{d}t$.

We can further simplify equation~(\ref{eq:eventrate}) by making use of the relation $h_k = \sum_j F_{kj} g_j$ derived above. It is then possible to combine the two matrices $G_{ij}$ and $H_{ij}$ into one matrix
\begin{equation}
D_{ij} = G_{ij} + \sum_k F_{kj} H_{ik} \; ,
\label{eq:Dij}
\end{equation}
such that
\begin{equation}
R_i = \sum_j D_{ij} g_j \;.
\end{equation}
The final expression is thus formally identical to the one used in~\cite{Feldstein:2014gza}. In other words, it is possible to obtain the same concise expression relating the binned event rates and the discretized velocity integral even if the differential cross section has a more complicated velocity dependence. The matrix $D_{ij}$ encodes all relevant experimental details and the assumed DM properties, but it is independent of the DM velocity distribution. 

Once the matrix $D_{ij}$ has been calculated, it is possible to calculate the predicted number of events for each bin for any given velocity integral parametrized by the coefficients $g_j$. These predictions can be used to calculate the likelihood for a specific choice of the $g_j$. Denoting the expected number of background events in each bin by $B_i$ and the total number of observed events by $N_i$, this likelihood is given by
\begin{equation}
-2 \log \mathcal{L} = 2 \sum \left[ R_i + B_i - N_i + N_i \log \frac{N_i}{R_i + B_i} \right] \; .
\label{eq:binnedL}
\end{equation}
Since all entries of $D_{ij}$ are positive, one can show that this likelihood is a convex function of the parameters $g_j$ defined on a convex domain and therefore any local minimum is guaranteed to be a global minimum. In other words, in spite of the large number of steps $N_\mathrm{s}$ it is possible to unambiguously minimise the likelihood with respect to the DM velocity distribution and thereby find the astrophysical parameters that give the best fit to a given set of observations.\footnote{We emphasize that there may in principle be several degenerate minima, i.e.\ different velocity distributions that yield the same likelihood. This does not pose a difficulty for our approach, as we are interested only in the minimum value of the likelihood and not in the underlying velocity distribution. We refer to~\cite{Feldstein:2014gza} for a more detailed discussion.}

The equations above generalise directly to the case of several experiments (labelled by $\alpha = 1, \ldots, N_\text{exp}$). In this case, the matrix $D^{(\alpha)}_{ij}$ must be calculated separately for each experiment.\footnote{Note that the number of bins and therefore the range of the index $i$ may be different for each experiment. However, the number of steps in $v_\text{min}$-space and hence the range of the index $j$ is the same for all experiments.} One can then calculate the likelihood $\mathcal{L}^{(\alpha)}$ for each experiment given the expected number of background events $B_i^{(\alpha)}$ and the total number of observed events $N_i^{(\alpha)}$. The total likelihood for given $g_j$ is then calculated by multiplying together the likelihoods for the individual experiments.

Additional technical details on how to calculate the matrices $G_{ij}$ and $H_{ij}$ are provided in the appendices. Appendix~\ref{ap:splitting} presents the method we use in order to implement equations (\ref{eq:Gij})--(\ref{eq:Dij}) in a numerically efficient way. In appendix~\ref{ap:highv} we discuss the problem of determining the appropriate ranges of $v_\text{min}$-space for DM interactions with non-standard velocity dependence.

\section{Studying generalised DM interactions}
\label{sec:generalised}

In the remainder of this paper we illustrate the possible applications of the framework developed above by addressing the following question: Given a set of experimental data, what information can be inferred on the coupling structure of DM without making any assumptions on its astrophysical distribution? To answer this question, we will generate sets of mock data for various DM models and then determine the goodness-of-fit for a number of different coupling structures. The resulting information can be used both to understand qualitatively whether certain coupling scenarios are distinguishable in principle as well as to quantify how strongly a specific coupling scenario is favoured or disfavoured by a given set of data.

In this section, we present the models that we consider for this study and illustrate the importance of halo-independent methods. Furthermore, we explain the quantitative method used to compare the different coupling structures. In section~\ref{sec:1exp} we then focus on the simplest case, where only a single experiment is sensitive to the interactions of DM. More complicated scenarios with more than one experiment are discussed in section~\ref{sec:2exp}.

\subsection{Models}
\label{sec:models}

We apply the halo-independent approach introduced in section~\ref{sec:haloindependent} to five different benchmark models of DM interactions: standard spin-independent and spin-dependent scattering, DM featuring an anapole or a magnetic dipole moment, as well as DM interacting via a ``dark magnetic dipole moment''. Our main reasoning for choosing these scenarios, which will be discussed in more detail below, is that they encompass the most relevant nuclear response functions and shapes of recoil spectra, and thus should constitute a representative set for describing the phenomenology of possible future direct detection data. Moreover, each of our benchmark scenarios can arise in well-motivated UV complete models of DM, making our choice also motivated from the model building perspective. We remark that our set of scenarios is similar to the one discussed in~\cite{Gresham:2014vja,Gluscevic:2015sqa}.

In the following, we define each of the benchmark models in terms of an effective DM-nucleon Lagrangian, expressed via the non-relativistic operators $\mathcal{O}_\lambda^\tau$~\cite{Fitzpatrick:2012ix,Catena:2015uua} (see also equation~(\ref{eq:Leff})). For the generation of mock data for future experiments, we assume in all scenarios a true DM mass $m_\text{DM} = 50\,$GeV, as well as a standard Maxwell-Boltzmann velocity distribution, with the most probable speed $v_0 = 220\,\text{km}/\text{s}$, the Galactic escape velocity $v_0 = 544\,\text{km}/\text{s}$ and the mean velocity of the Earth $v_\text{E} = 230\,\text{km}/\text{s}$. Furthermore, we fix the normalization of the interaction strength (e.g.~the total scattering cross section in the standard SI/SD case) such that it complies with all current direct detection bounds. Concretely, we consider the experiments LUX~\cite{Akerib:2015rjg}, SuperCDMS~\cite{Agnese:2014aze}, SIMPLE~\cite{Felizardo:2011uw}, COUPP~\cite{Behnke:2012ys}, PICASSO~\cite{Archambault:2012pm} and PICO-2L~\cite{Amole:2016pye}, all of them implemented as in~\cite{Catena:2016hoj}, and choose the normalization of the interaction rate in each benchmark model to be a factor of two below the most sensitive exclusion bound.

\subsubsection*{Standard spin-independent and spin-dependent scattering (SI/SD)}

A large variety of DM models lead to the standard spin-independent and/or spin-dependent interaction, which are described by the (relativistic) effective DM-nucleon Lagrangians\linebreak$f_N^{\text{(SI)}} \bar \chi \chi \, \bar N N$ and $ f_N^{\text{(SD)}} \bar \chi \gamma^\mu \gamma^5 \chi \, \bar N \gamma_\mu \gamma^5 N$, respectively. In the notation of~\cite{Catena:2015uua}, the resulting non-relativistic Lagrangians read
\begin{align}
\mathcal{L}_\text{eff} = \sum_{N=p,n} f_N^{\text{(SI)}} \mathcal{O}_1^{(N)} \quad \text{and} \quad \mathcal{L}_\text{eff} = -4 \sum_{N=p,n} f_N^{\text{(SD)}} \mathcal{O}_4^{(N)} \,,
\end{align}
respectively.\footnote{We refer to appendix~\ref{ap:non-relativistic} for the conversion of such a linear combination of non-relativistic operators into a scattering cross section $\text{d}\sigma/\text{d}E_\mathrm{R}$.} Concretely, for the spin-independent case, we will consider a benchmark scenario with fixed neutron-to-proton coupling ratio $f_n/f_p=1$, corresponding e.g.~to Higgs exchange, with an absolute strength given by $\sigma_p^{\text{(SI)}} = 5 \cdot 10^{-46}\,\text{cm}^2$. For spin-dependent scattering we consider a model with $f_n^{\text{(SD)}}/f_p^{\text{(SD)}}=-1$ (motivated by a scenario of Majorana DM interacting with nucleons via the Standard Model $Z$-boson) with a total cross section $\sigma_p^{\text{(SD)}} = 1.2 \cdot 10^{-40}\,\text{cm}^2$. Note that while we restrict ourselves to specific values of $f_n/f_p$ for the purpose of generating the mock data, we will in some cases allow arbitrary neutron-to-proton coupling ratios for the halo-independent fits that we perform.

\subsubsection*{Anapole moment (AM)}

Next, we consider DM interacting dominantly via an anapole moment, corresponding to the DM-photon interaction $\mathcal{A}\,  \bar \chi \gamma^\mu \gamma^5 \chi \, \partial^\nu F_{\mu \nu}$, which constitutes the only possible electromagnetic moment of a Majorana fermion; see e.g.~\cite{Chang:2014tea} for a concrete realization. The coupling of the anapole moment to the nucleon electromagnetic current then leads to~\cite{Gresham:2014vja}
\begin{align}
\mathcal{L}_\text{eff} = 2 \mathcal{A} e \sum_{N=p,n} \left( Q_N \mathcal{O}_8^{(N)} + \widetilde{\mu}_N \mathcal{O}_9^{(N)} \right) \,,
\end{align}
where $e$ is the electric charge, $Q_p = 1$ and $Q_n=0$ is the charge of the proton and neutron in units of $|e|$, respectively, and $\widetilde{\mu}_N$ is the magnetic moment of the nucleon $N$ in units of the nuclear magneton $e/(2m_p)$. This interaction structure gives rise to a scattering cross section with a non-standard dependence on $v^2$ and $q^2$: it contains a spin-independent part with an additional factor of $v_\perp^2 \equiv v^2-q^2/(4 \mu_T^2)$, as well as a contribution rising as $q^2$, which is enhanced for nuclei with a significant magnetic dipole moment. For the generation of the mock data, we again assume $m_\text{DM}=50\,$GeV, as well as an anapole moment \mbox{$\mathcal{A} = 3.5 \cdot 10^{-5} \,\text{GeV}^{-2}$}, compatible with all current direct detection bounds.

\subsubsection*{Magnetic dipole moment (MDM)}

Dirac DM can interact with photons via a magnetic dipole moment given by $\mathcal{D}_\text{magn}\,  \bar \chi \sigma^{\mu \nu} \chi F_{\mu \nu}$ (there is also an equivalent operator for complex scalar DM). Scattering induced by the magnetic dipole moment of DM is known to be the leading contribution to the total interaction rate e.g.~in scenarios where a Dirac DM particle couples to a Standard model lepton and a heavy scalar mediator via a Yukawa interaction~\cite{Chang:2014tea,Ibarra:2015fqa}. The effective non-relativistic DM-nucleon Lagrangian in this scenario is given by~\cite{Gresham:2014vja}
\begin{align}
\mathcal{L}_\text{eff} = \frac{2 \mathcal{D}_\text{magn} e}{q^2} \sum_{N=p,n} \left[ Q_N \left( m_N \mathcal{O}_5^{(N)} - \frac{q^2}{4 m_\chi} \mathcal{O}_1^{(N)} \right) + \widetilde{\mu}_N \left( m_N \mathcal{O}_6^{(N)} - \frac{q^2}{m_N} \mathcal{O}_4^{(N)} \right) \right] \,.
\label{eq:Leff_MD}
\end{align}
Compared to the standard spin-independent or spin-dependent scattering, this interaction structure leads to a differential scattering cross section which is enhanced by $1/E_\text{R}$ at small recoil energies, implying a steeper recoil spectrum. For our benchmark scenario, we choose $\mathcal{D}_\text{magn} = 9.3 \cdot 10^{-7} \,e\, \text{fm}$.

Similar to the magnetic dipole moment, DM could also interact with nucleons via an electric dipole moment (which however would require significant CP violation in the dark sector)~\cite{Barger:2010gv,Banks:2010eh}. We have found that for the purpose of this work, magnetic and electric dipole DM are essentially equivalent from a phenomenological perspective, and hence for simplicity we only include the former scenario in our list of benchmark models.

\subsubsection*{Dark magnetic dipole moment (DMDM)}

Lastly, we consider the case of a ``dark magnetic dipole moment'', which is obtained by replacing $1/q^2 \rightarrow 1/M^2$ in the MDM scenario given by equation~(\ref{eq:Leff_MD})~\cite{Fitzpatrick:2010br}. Physically, this corresponds to a situation in which DM interacts via a dipole operator with a new heavy gauge boson of mass $M$, as opposed to the MDM scenario involving the exchange of a  photon. Redefining the overall normalization of the interaction strength as $\mathcal{D}_\text{DMDM}$ (which has units GeV$^{-3}$), we obtain
\begin{align}
\mathcal{L}_\text{eff} = 2 \mathcal{D}_\text{DMDM} \sum_{N=p,n} \left[ Q_N \left( m_N \mathcal{O}_5^{(N)} - \frac{q^2}{4 m_\chi} \mathcal{O}_1^{(N)} \right) + \widetilde{\mu}_N \left( m_N \mathcal{O}_6^{(N)} - \frac{q^2}{m_N} \mathcal{O}_4^{(N)} \right) \right] \,.
\end{align}
Phenomenologically, this model has the interesting feature that it leads to a recoil spectrum rising like $q^2$ at small momentum transfers, strikingly different from the standard spin-independent or spin-dependent scenarios. For the generation of the mock data corresponding to this benchmark model, we assume $\mathcal{D}_\text{DMDM} = 1.6 \cdot 10^{-4} \, \text{GeV}^{-3}$.

\subsection{Mapping to $v_\text{min}$-space}
\label{sec:vminspace}

For a given velocity integral the models introduced above make very different predictions for the number of expected events in different experiments as well as for the shape of the individual recoil spectra. Thus, one would expect that they can be easily distinguished once a sizeable number of DM scattering events have been observed. Such a reconstruction of the DM properties, however, is often only possible if a specific form of the velocity integral is assumed. Unfortunately, incorrect assumptions on the velocity integral may strongly bias the reconstruction and may even lead to the exclusion of the correct model of DM~\cite{Peter:2011eu,Pato:2012fw,Feldstein:2014gza}.

The halo-independent method introduced in section~\ref{sec:haloindependent} does not require such assumptions and therefore avoids false exclusions. A direct consequence is that it becomes harder in such an approach to distinguish different models of DM. Whenever changes in the velocity distribution can compensate for changes in the particle physics properties of DM, it is conceivable that a good fit to the data can be obtained even for an incorrect model of DM. It is therefore of great importance to understand what experimental set-up is necessary to ensure that different models of DM can be distinguished even when allowing for arbitrary velocity distributions.

A convenient way to illustrate these issues is to map experimental data into $v_\text{min}$-space~\cite{Fox:2010bz,Frandsen:2011gi,Gondolo:2012rs}. Here we briefly review the general idea of this mapping and then use it to discuss a few illustrative examples for the importance of halo-independent methods. For this purpose let us consider an experiment that has observed a number of DM scattering events, allowing it to infer the differential event rate $\mathrm{d}R/\mathrm{d}E_\mathrm{R}$ at a certain recoil energy $E_0$. Under the assumption of a specific DM model, this measurement can be used to infer the value of the velocity integral $g(v_\text{min,0})$ for $v_\text{min,0} = v_\text{min}(E_0)$. For example, for SI interactions with $f_n/f_p=1$, one finds
\begin{equation}
\label{eq:usualSI}
\frac{\text{d}\sigma}{\text{d} E_\text{R}} = A^2 \, F^2(E_{\text{R}}) \frac{m_T \, \sigma_n}{2 \, \mu_{n\chi}^2 \, v^2}\;,
\end{equation}
where $A$ is the mass number of the nucleus, $F(E_\text{R})$ is the standard SI form factor and $\sigma_n$ is the DM-neutron scattering cross section at zero momentum transfer. Substituting this expression into equation~(\ref{eq:dRdEsimple}), one finds
\begin{equation}
\frac{\text{d}R}{\text{d}E_\text{R}} = \frac{\rho \, A^2 \, F^2(E_{\text{R}}) \, \sigma_n}{2 \, \mu_{n\chi}^2 \, m_\chi} g(v_\text{min}) \; .
\label{eq:dRdESI}
\end{equation}

To simplify this expression, we absorb all quantities that do not depend on the target material into the \emph{rescaled velocity integral}
\begin{equation}
\tilde{g}(v_\text{min}) \equiv \frac{\rho \, \sigma_n}{m_\chi} g(v_\text{min}) \; .
\end{equation}
The rescaled velocity integral can now be calculated from the measured differential event rate by inverting eq.~(\ref{eq:dRdESI}):
\begin{equation}
\tilde{g}(v_\text{min,0}) = \frac{2 \, \mu_{n\chi}^2}{A^2 \, F^2(E_{\text{R}})} \frac{\text{d}R}{\text{d}E_\text{R}}(E_0)\; .
\end{equation}
This translation can be performed for further measurements of the differential event rate at different recoil energies and for measurements made by additional experiments.\footnote{We emphasise that in practice both the measurement of $\mathrm{d}R/\mathrm{d}E_\mathrm{R}$ and of the corresponding recoil energy $E_0$ will have some experimental error due to limited statistics and finite energy resolution of the detector. Although we do not discuss these complications in detail here, they are taken into account in the translations that we show. For further details, we refer to~\cite{Gondolo:2012rs}. We furthermore note that the mapping to $v_\text{min}$-space becomes more complicated if the differential cross section has a non-trivial dependence on the DM velocity~\cite{DelNobile:2013cva}. To avoid this complication, we only consider the mapping for SI interactions.} Since all experiments probe the same velocity integral, it should be possible to fit the inferred values for $\tilde{g}(v_\text{min})$ with a single monotonically decreasing function. If no good fit can be found, the assumed model of DM can be ruled out in a halo-independent way.

\begin{figure}
\centering
\includegraphics[width=0.45\textwidth]{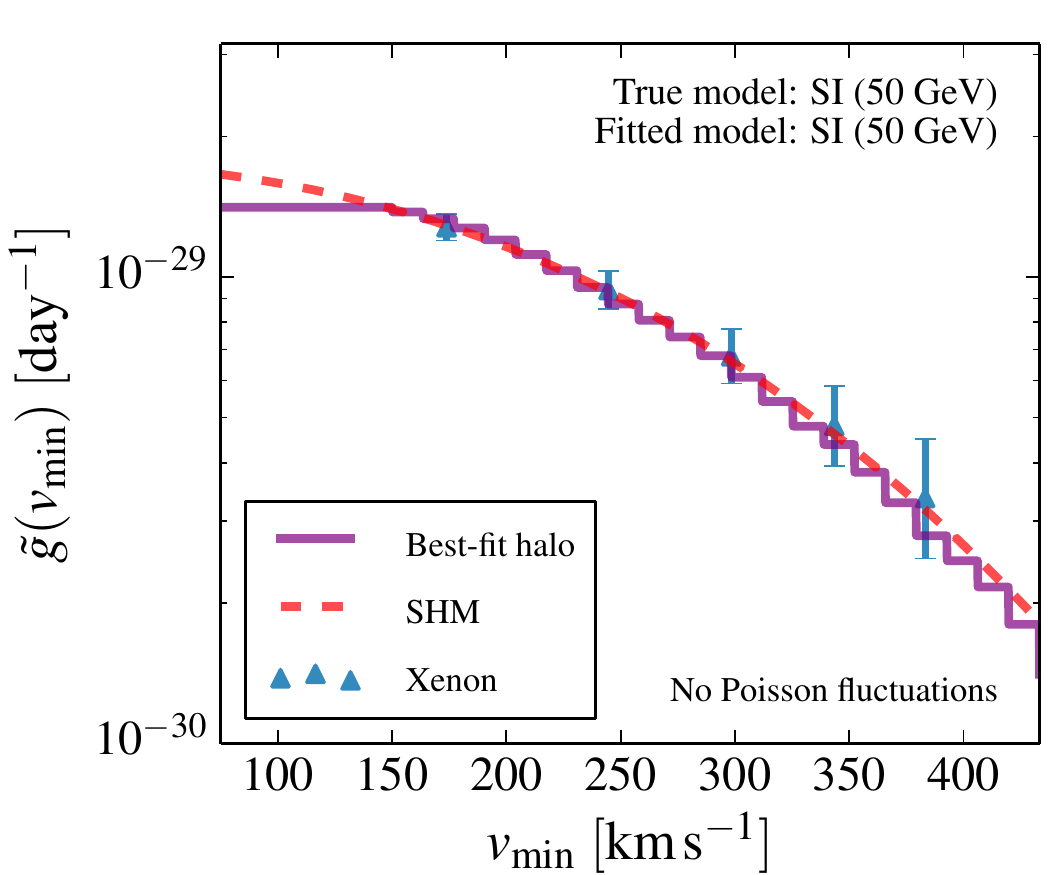}\qquad
\includegraphics[width=0.45\textwidth]{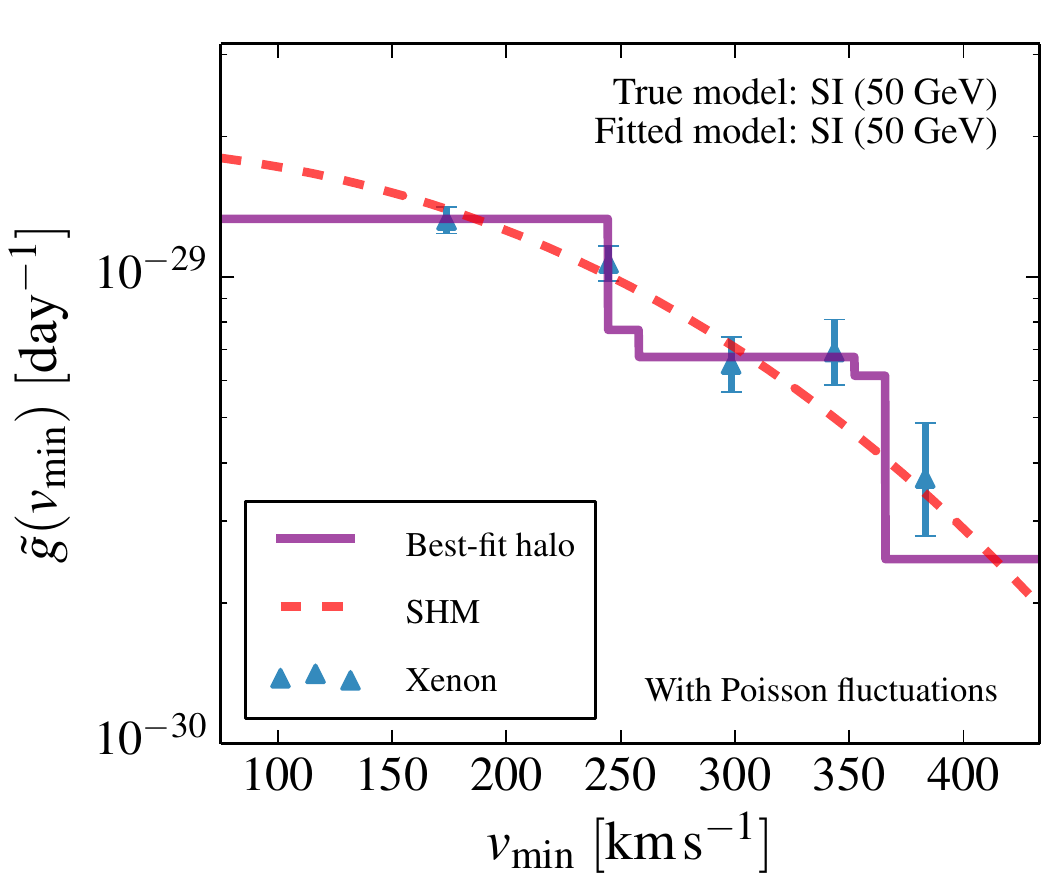}
\includegraphics[width=0.45\textwidth]{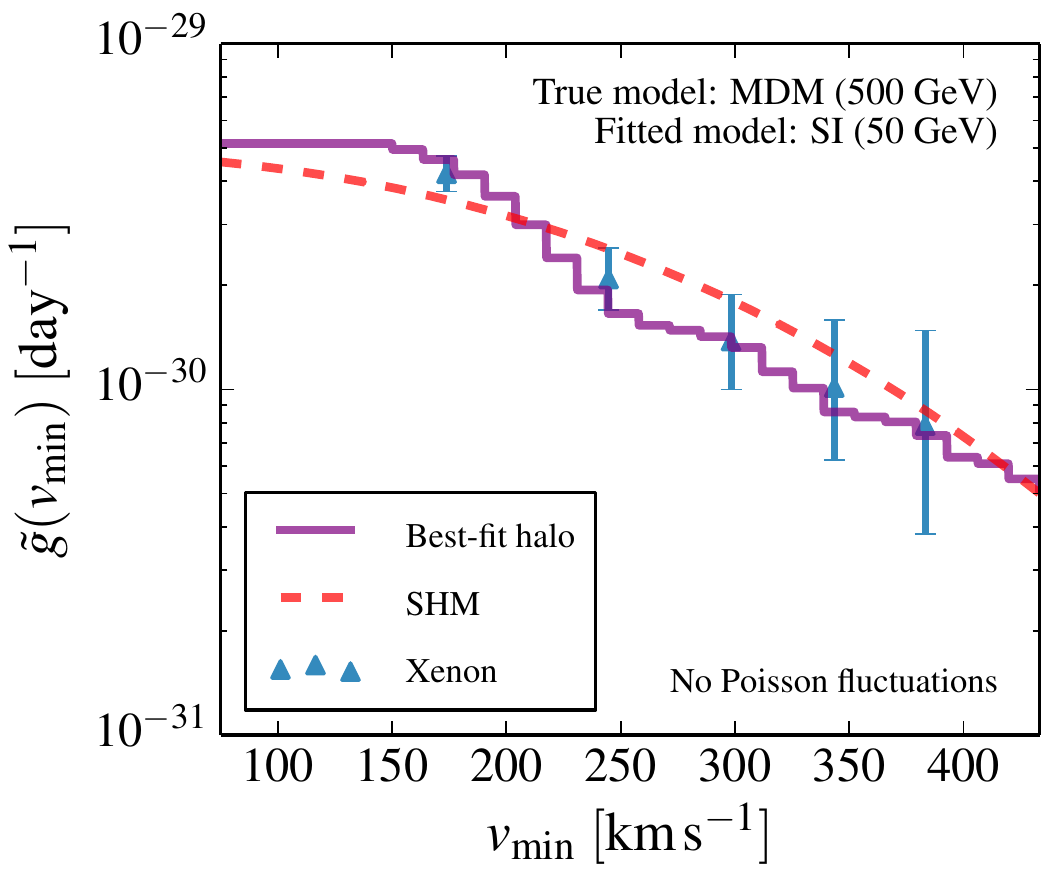}\qquad
\includegraphics[width=0.45\textwidth]{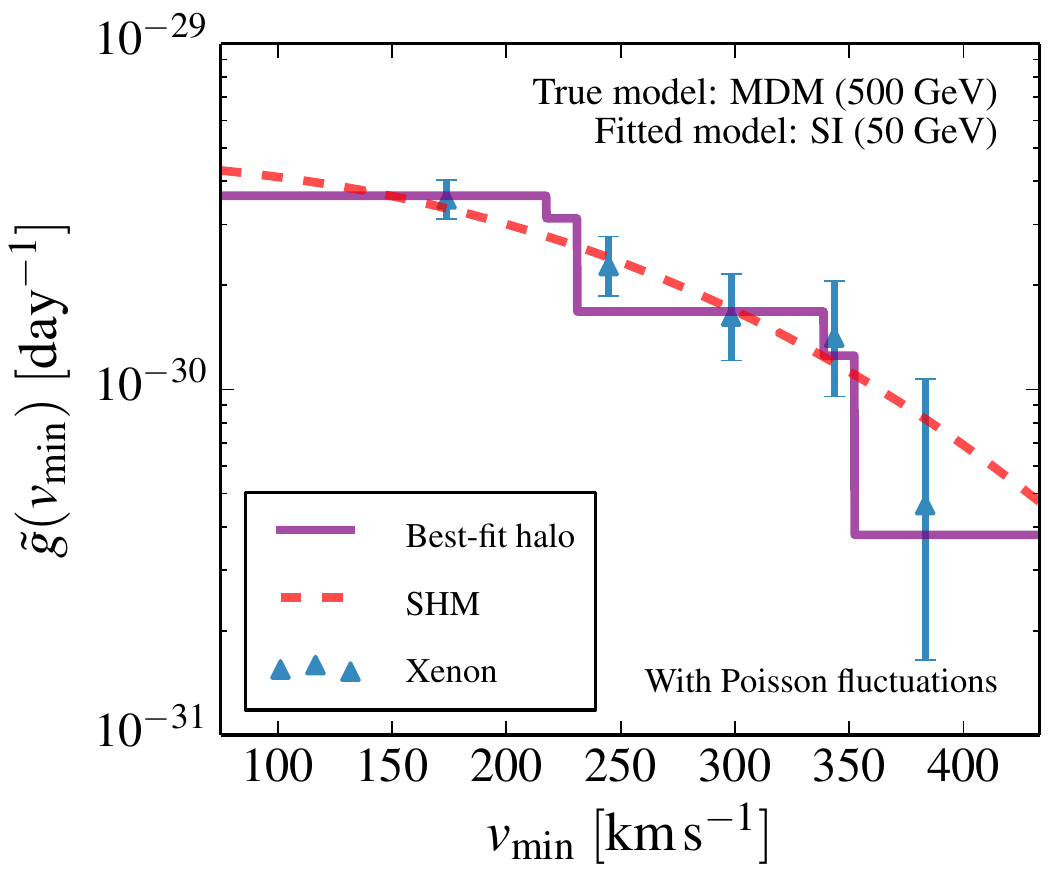}
\caption{Reconstruction of the velocity integral in comparison with the underlying mock data mapped to \mbox{$v_\text{min}$-space}. In the top row the same model (SI interactions) has been used to generate the data and to reconstruct the velocity integral, while in the bottom row an incorrect model (MDM interactions) has been assumed for the reconstruction. In the right column Poisson fluctuations have been applied to the mock data to illustrate their effect on the best-fit velocity integral. Note that the error bars on the data points are shown for illustration only and are not used directly for the calculation of the best-fit velocity integral, which is determined from the binned event rates as described in section~\ref{sec:haloindependent}.}
\label{fig:vminspace1}
\end{figure}

In figure~\ref{fig:vminspace1} we illustrate this mapping using mock data from a single experiment based on liquid xenon (Xe), see section~\ref{sec:1exp} for details. The top-left panel shows the simplest case, where we use the same model, namely SI interactions with $m_\chi = 50\:\text{GeV}$, to generate the mock data and to perform the mapping to $v_\text{min}$-space. In the absence of Poisson fluctuations, one can see that the inferred values of $\tilde{g}(v_\text{min})$ agree perfectly with the SHM used to generate the mock data (indicated by the red dashed line).\footnote{To study the case with no Poisson fluctuations we employ our definition of the likelihood in eq.~(\ref{eq:binnedL}) also for non-integer values of $N_i$.} The best-fit velocity integral for $N_\mathrm{s} = 30$ is indicated by the purple solid line. As expected, it agrees well with the SHM.

In the top-right panel, we have included Poisson fluctuations in the data. As a result, the best-fit velocity integral now differs from the SHM, because it attempts to follow the random fluctuations in the data. The important observation is that even the best-fit velocity integral cannot necessarily achieve a perfect fit to the data, because it is required to be monotonically decreasing and can therefore not always follow upward fluctuations at high energies.

In the bottom row we consider a different situation and use a different model to generate the mock data, namely MDM interactions with $m_\chi = 500\:\text{GeV}$, but still use SI interactions with $m_\chi = 50\:\text{GeV}$ for the mapping to $v_\text{min}$-space. The different shape of the recoil spectrum for MDM interactions implies that the SHM no longer gives a good fit to the data. Nevertheless, by considering arbitrary velocity integrals it is still possible to obtain a good fit. In other words, it is not possible in this example to rule out SI interactions in a halo-independent way even if DM scattering actually arises from MDM interactions.

\begin{figure}
\centering
\includegraphics[width=0.45\textwidth]{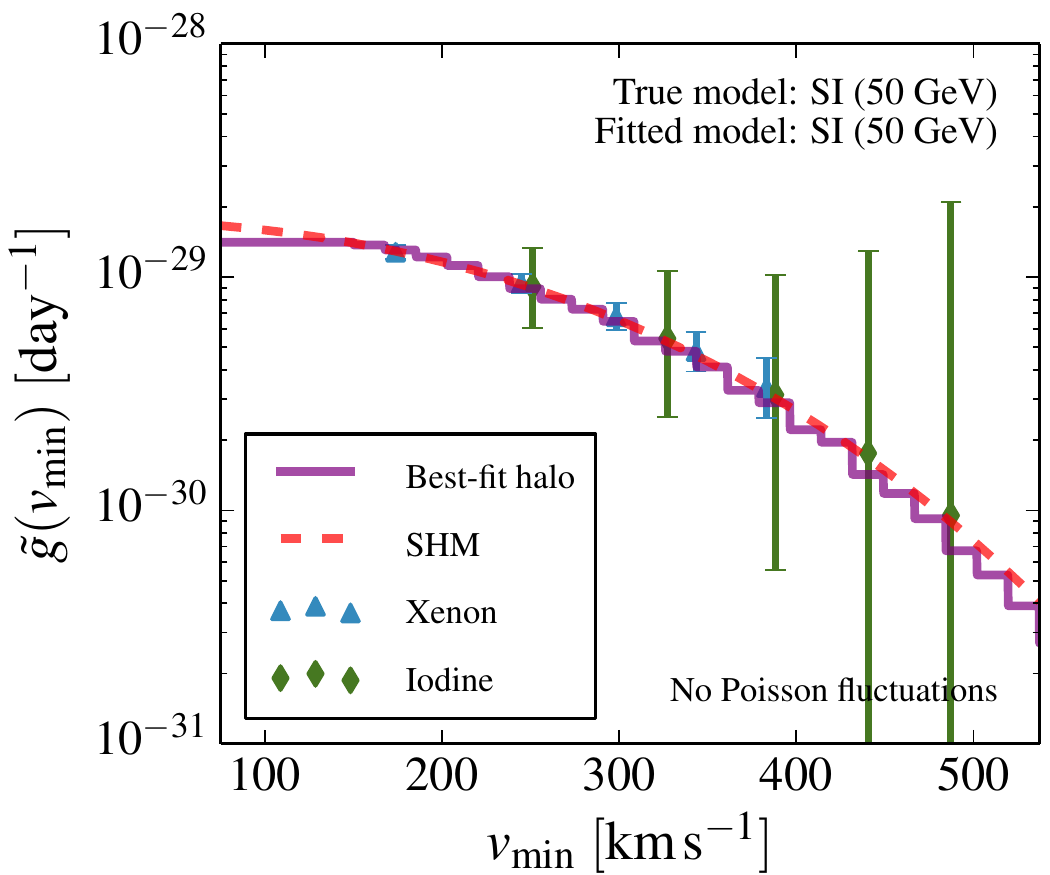}\qquad
\includegraphics[width=0.45\textwidth]{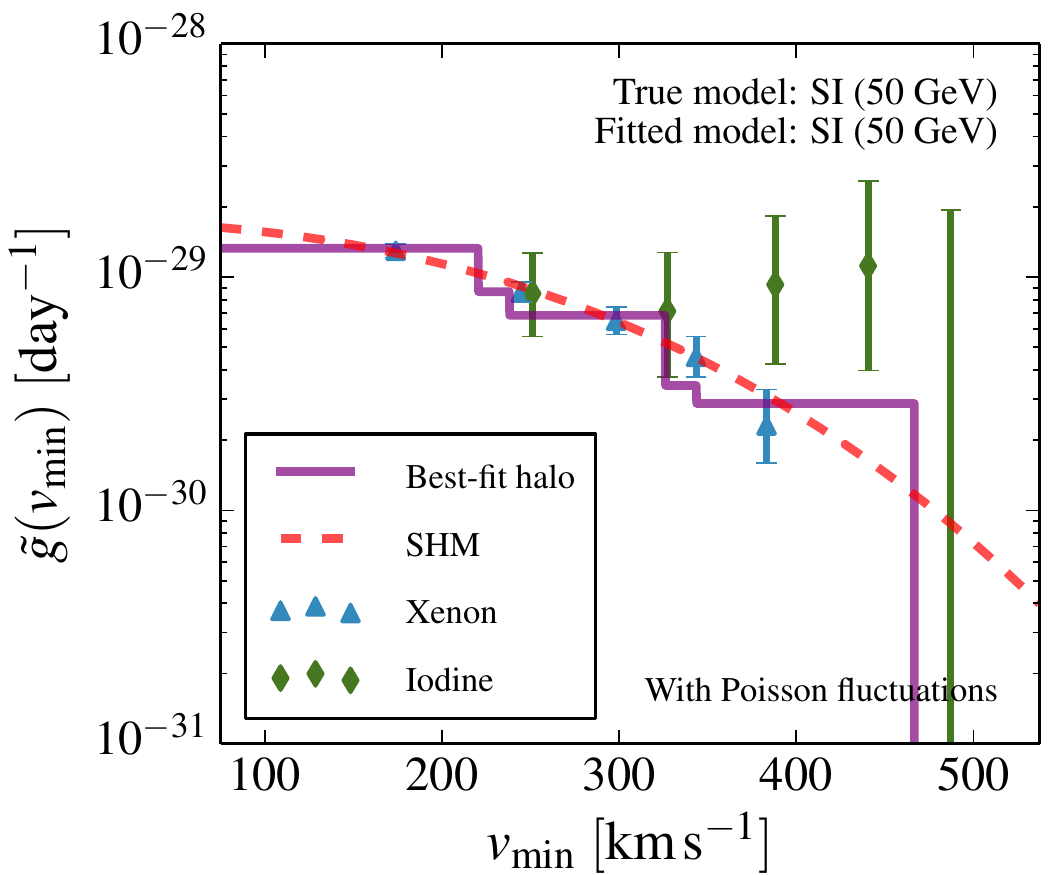}
\includegraphics[width=0.45\textwidth]{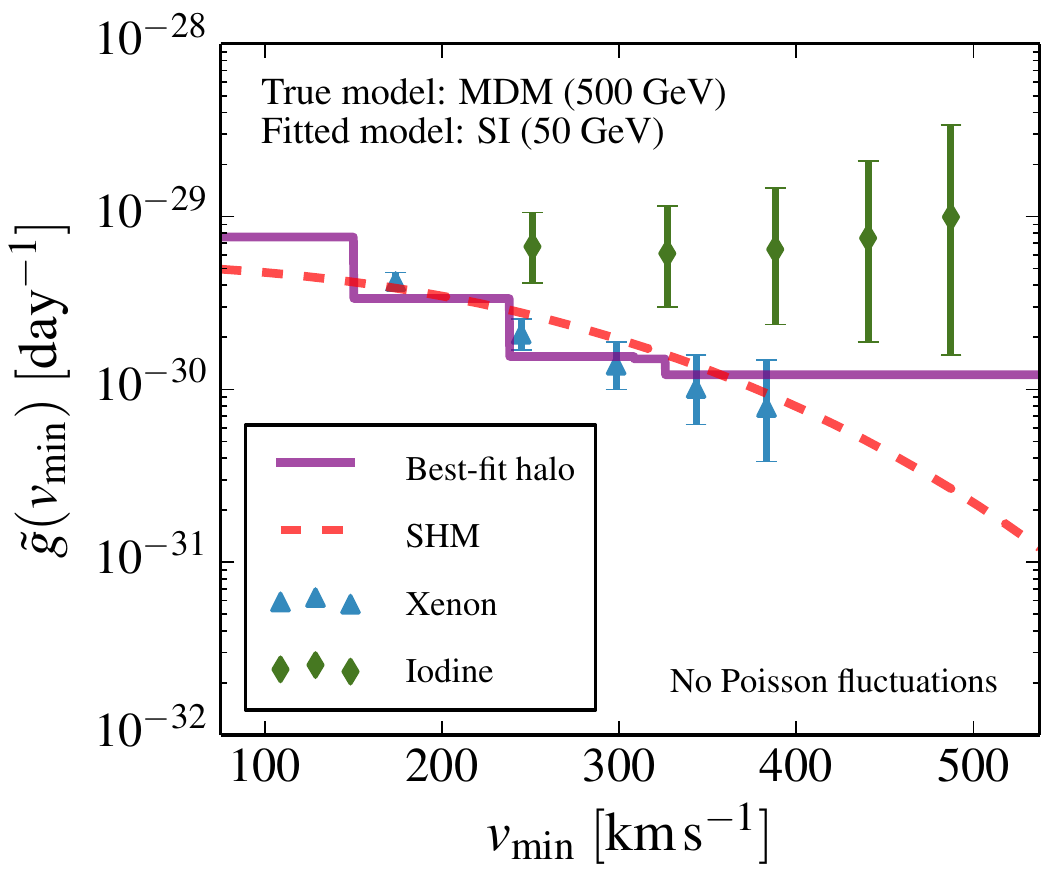}\qquad
\includegraphics[width=0.45\textwidth]{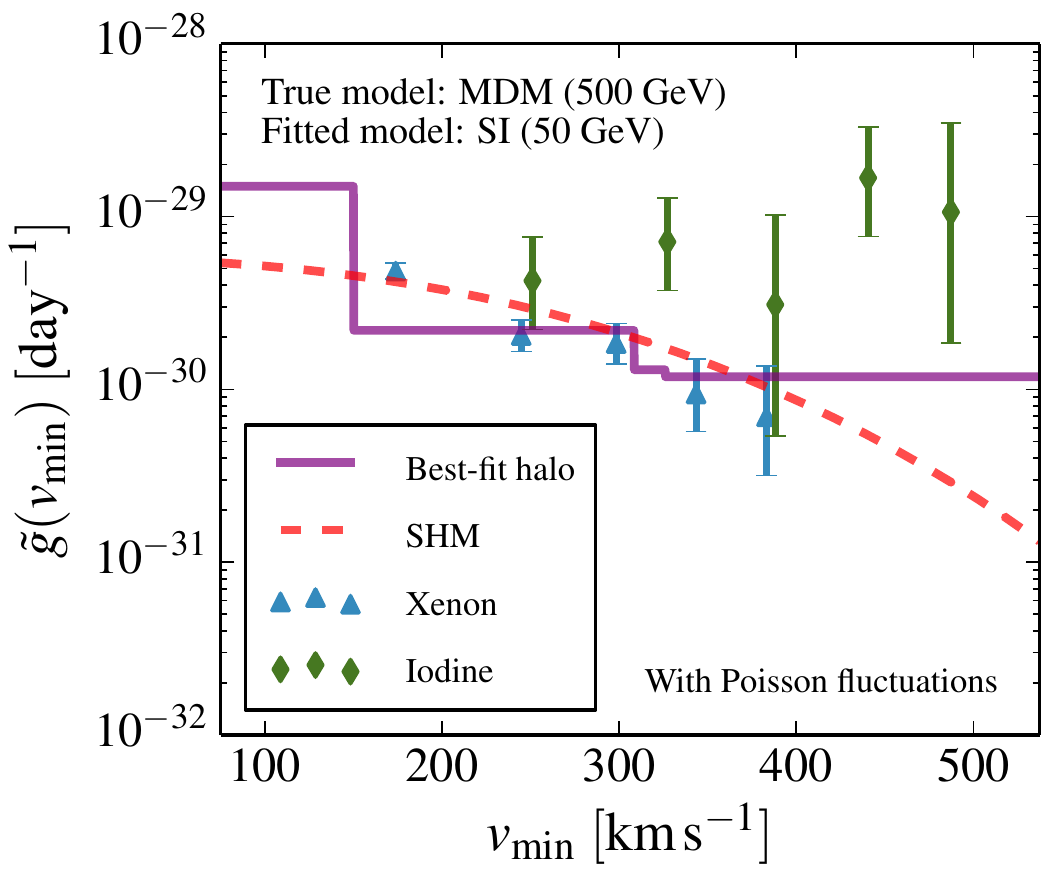}
\caption{Same as figure~\ref{fig:vminspace1}, but using the combined information from a xenon experiment and an iodine experiment for the reconstruction.}
\label{fig:vminspace2}
\end{figure}

Clearly, this situation changes significantly if two experiments are considered (see section~\ref{sec:2exp} for details), because the two experiments will provide independent measurements of the rescaled velocity integral at the same values of $v_\text{min}$. This is illustrated in figure~\ref{fig:vminspace2}. In the top row we again consider the case that the true model is identical to the fitted model. As expected, the measurements from both experiments are in good agreement with each other and a consistent fit can be found, even though the fit will be less than perfect in the presence of Poisson fluctuations. Using MDM interactions to generate the mock data, we however find strong incompatibility between the inferred values of $\tilde{g}(v_\text{min})$ from the two experiments. The reason is essentially that, compared to xenon targets, iodine targets are more sensitive to MDM interactions than to SI interactions.

Obviously, it is difficult to quantify these statements using the information from figure~\ref{fig:vminspace2} alone. For example, once Poisson fluctuations are included it becomes less trivial to distinguish the top row from the bottom row. In particular, it is difficult from these plots to appreciate how sensitive experimental predictions are to variations in the velocity integral. A discussion of how to construct confidence bands in $v_\text{min}$-space can be found in~\cite{Gelmini:2015voa}. Here we take a different approach and focus entirely on the likelihood constructed from the binned event rates in energy space, which is given in eq.~(\ref{eq:binnedL}). We now discuss how one can use this approach to make rigorous quantitative statements about the incompatibility of different measurements and to conclude whether or not the interactions under consideration can be distinguished.

\subsection{Goodness-of-fit estimates}
\label{sec:gof}

For each of the models discussed above, and given a value of the DM mass $m_\chi$ and the DM velocity distribution $f(v)$, we can calculate the predicted number of events in a given experiment. 
Once we have calculated the predicted number of events in each bin, we can apply Poisson fluctuations to generate mock data. Each of the models from above can then be fitted to the mock data in order to quantify the likelihood that the data corresponds to this particular model, using the definition of $\mathcal{L}$ given in equation~(\ref{eq:binnedL}). If the \emph{fitted model} is identical to the \emph{true model} (assumed for generating the mock data), we generally expect to find a good fit, i.e.\ a likelihood within the range expected by random fluctuations in the data. If the fitted model is different from the true model, the likelihood may be much smaller, depending on how well true model and fitted model can be distinguished by the experimental set-up.

For a given set of mock data and a chosen model, we can determine the value of the DM mass that maximises the likelihood (i.e.\ that minimises $-2 \log \mathcal{L}$). Ideally, the resulting minimum value of $-2 \log \mathcal{L}$ (called $x_0$) could simply be compared to a $\chi^2$ distribution with the appropriate number of degrees of freedom in order to quantify the goodness-of-fit of the best-fit point. Unfortunately, as discussed in~\cite{Feldstein:2014gza}, this simple approach does not work in our context. Although the distribution of $x_0$ does approximately follow a $\chi^2$ distribution (when considering random fluctuations in the underlying data), we cannot easily determine the appropriate number of degrees of freedom. The reason is that finding the optimum velocity integral typically requires a very large number of fitted parameters, many of which simply saturate the boundary condition from the monotonicity requirement without actually improving the fit to the data. In other words, even if the number of fitted parameters significantly exceeds the number of bins, we typically do not obtain a perfect fit, although the naive number of degrees of freedom should be zero.

While it is thus not possible to determine analytically the expected distribution for $x_0$, this distribution can be extracted from Monte Carlo simulations~\cite{Feldstein:2014ufa}. For this purpose, we can simply take the best-fit point of the fitted model as the basis for generating new sets of mock data and then repeat the same fitting procedure as above. If the typical values for $-2 \log \mathcal{L}$ determined in this second step (called $x_1$) are comparable to the one found in the first step above, the conclusion would be that the assumed model gives a good fit to the data. If, on the other hand, the first step gave a much larger value for $x_0$ than what is typically found when taking the best-fit point to generate new sets of mock data, the goodness-of-fit is bad. More precisely, if $\zeta(x_1)$ is the (normalised) distribution of $x_1$ obtained from Monte Carlo simulations of the best-fit point, we can define the $p$-value for the assumed model as
\begin{equation}
 p = \int_{x_0}^\infty \zeta(x_1) \mathrm{d}x_1 \; .
\end{equation}

The procedure outlined above is straight-forward to carry out in the case of a single set of experimental data. Lacking a conclusive DM signal, however, we are interested in the typical performance of future experiments. Such a study requires a large number of mock data sets. We therefore adopt the following procedure:
\begin{enumerate}
 \item For a given choice of the \emph{true model} (used to generate the mock data) and of the \emph{fitted model} (used to fit the mock data), we generate mock data sets and determine the best-fit DM mass, as well as the corresponding value of $x_0$ and the predicted number of events in each bin for the fitted model.
 \item Once we have determined the best-fit models for a large number of mock data sets, we find a \emph{typical prediction} of the fitted model by taking the median of the predicted number of events in each bin.
 \item We then take these typical predictions to generate more mock data sets and once again repeat the fitting procedure from step 1 to determine $x_1$.
\end{enumerate}
For each combination of true model and fitted model, we end up with two distributions: the distribution of $x_0$, corresponding to mock data generated from the true model, and the distribution of $x_1$, corresponding to mock data generated from the typical predictions of the fitted model.

\begin{figure}
\centering
\includegraphics[width=0.45\textwidth]{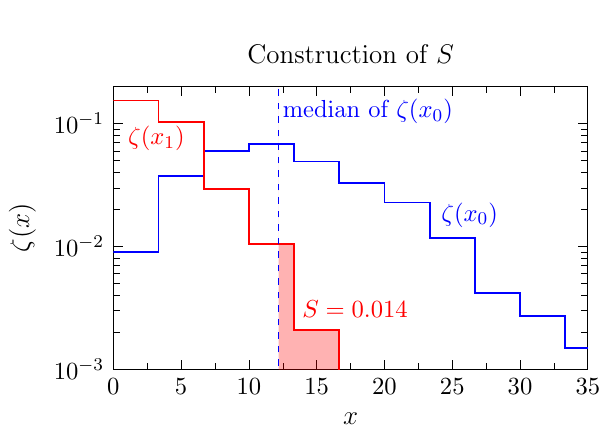}\qquad
\includegraphics[width=0.45\textwidth]{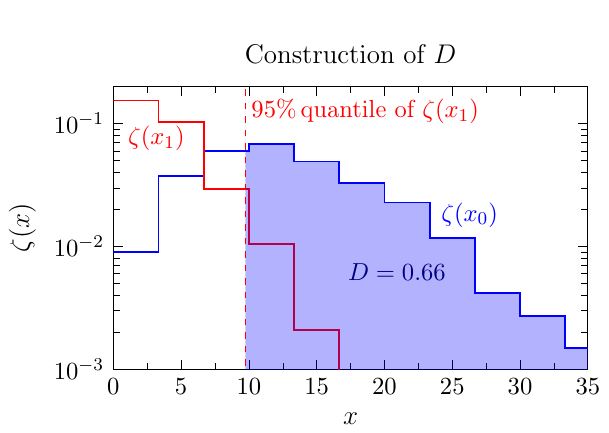}
\caption{Illustration of our definition of the similarity $S$ (left panel) and the distinguishability $D$ (right panel). See the text for details.}
\label{fig:illustration}
\end{figure}

If the two distribution are very similar, the conclusion is that true model and fitted model cannot be easily distinguished with the given experimental set-up and the chosen test statistic.\footnote{We note that it is still possible that the distributions of $x_0$ and $x_1$ would look different for a different test statistic. In particular, using an unbinned instead of a binned likelihood would retain additional information which may lead to stronger discrimination power. The use of such unbinned likelihoods is however computationally very expensive and therefore left for future work.} If the two distributions are different, we can use the results to address a number of interesting questions. First, what is the $p$-value corresponding to a typical value for $x_0$ (taking for example the median of $x_0$ across all mock data sets)? And second, in what fraction of the simulations is the $p$-value smaller than $5\%$, such that we would be able to exclude the fitted model at 95\% confidence level? We shall refer to the former number as similarity ($S$) and the latter number as the distinguishability ($D$) of the two models. The two quantities are illustrated in figure~\ref{fig:illustration}. For completely indistinguishable models (in particular the case that true model and fitted model are identical), one would obtain $S \approx 50\%$ and $D \approx 5\%$. Conversely, if $S < 5\%$ or $D > 50\%$ a typical realisation of the true model would allow to exclude the fitted model with at least $95\%$ confidence level (C.L.).

Clearly it is an approximation to calculate the distribution $\zeta(x_1)$ only for a typical prediction of the fitted model (defined in step 2 of the procedure presented above), rather than for each set of mock data separately. In principle, one should perform a separate Monte Carlo simulation for each best-fit point in order to determine the $p$-value of each fitted model. One would then obtain a distribution of $p$-values, which can be used to determine the similarity (i.e.\ median $p$-value of the fitted model) and the distinguishability (i.e.\ the fraction of fitted models with a $p$-value smaller than 5\%).

\begin{figure}[t]
\centering
\includegraphics[width=0.62\textwidth]{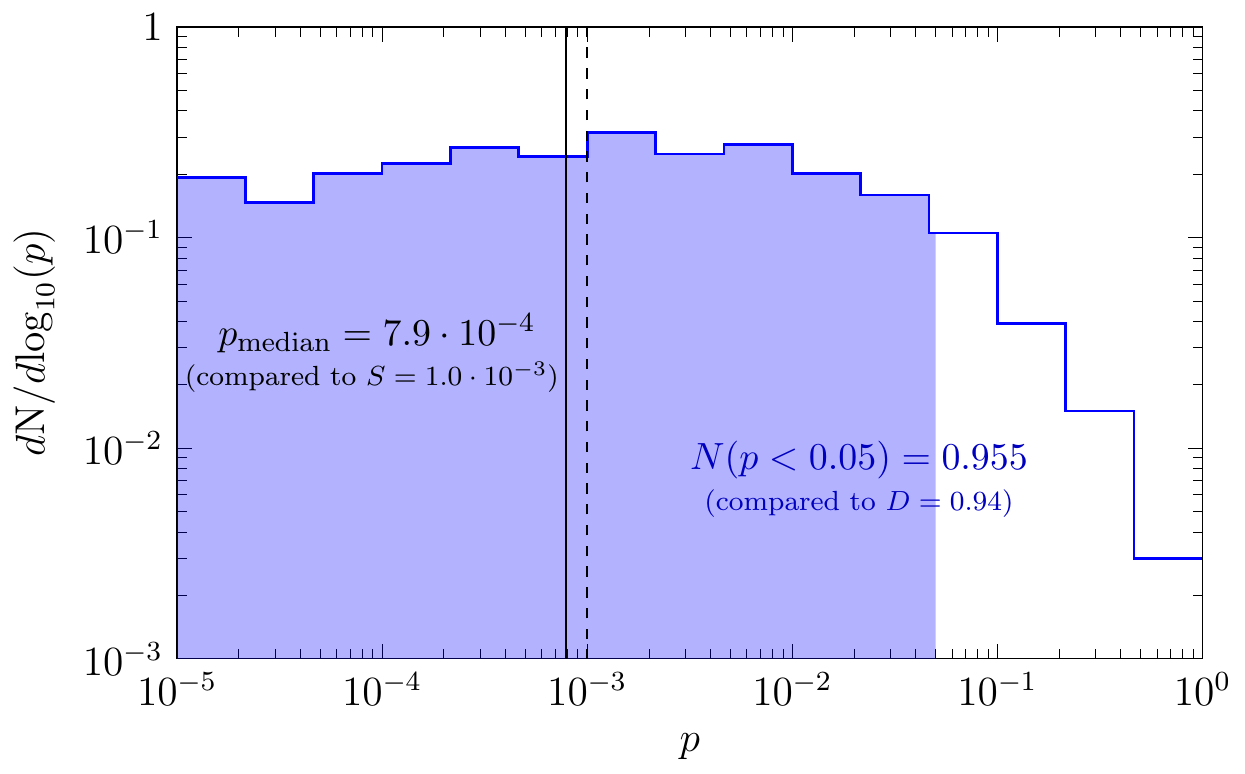}
\caption{Distribution of $p$-values obtained directly via Monte Carlo simulations, using for concreteness the true model DMDM and the fitted model SI. As indicated in the figure and explained in the text, in this example the similarity $S$ and the distinguishability $D$ obtained via our approximate method using a \emph{typical prediction} of the fitted model is in good agreement with the direct evaluation of these quantities from the distribution of $p$-values.}
\label{fig:pvalues}
\end{figure}

While it would be unfeasible to perform a separate Monte Carlo simulation for each realisation of the mock data and each combination of true model and fitted model, we have performed one such study for the case that the true model is DMDM and the fitted model is SI, employing mock data for a xenon-based experiment (see the following section for more details). The resulting distribution of $p$-values is shown in figure~\ref{fig:pvalues}. The median $p$-value is found to be $0.079\%$, which agrees well with the value $S = 0.1\%$ inferred from the procedure defined above within the uncertainties resulting from limited Monte Carlo statistics. Conversely, the fraction of simulations with a $p$-value below $5\%$ is $95.5\%$, compared to the value $D = 94\%$ obtained via the approximate method. The conclusion is that the approximation to use a typical prediction rather than each individual prediction is sufficient, because the spread of the predictions from the best-fit points is not too large.

\section{Constraining DM coupling structures with a single experiment}
\label{sec:1exp}

In this section we consider a single experiment, namely a liquid-xenon-based ton-scale detector similar to XENON1T~\cite{Aprile:2015uzo}. Details on the assumed exposure, energy window and detector resolution are given in table~\ref{tab:experiments}. We assume the contribution from backgrounds to be negligible.\footnote{In order to achieve background-free data, a detector based on liquid xenon typically needs to reject nuclear recoils that look too similar to electron recoils, leading to a reduction in the total exposure. We do not consider this effect, nor the potential background from solar neutrinos.} Given current constraints on the interactions of DM, we can expect up to 200 DM scattering events in such an experiment. We will see that even in this minimal case there is some halo-independent sensitivity to the coupling structure of DM.

\begin{table*}[t]
\begin{center}
\begin{tabular}{ccccc}
\hline \hline
Target & Exposure [kg yr] & Energy range [keV] & Number of bins & Energy resolution [keV] \\
\hline \hline
Xe & 2000 & 4--50 & 7 & 0.6 $\sqrt{E_\mathrm{R} / 1\:\text{keV}}$ \\
Ge & 200 & 0.3--50 & 9 & 0.06 \\
I & 100 & 10--100 & 9 & 0.15 $\sqrt{E_\mathrm{R} / 1\:\text{keV}}$\\
\hline\hline
\end{tabular}
\caption{Specification of the assumed performances of the future direct detection experiments employed in this work. We assume uniform acceptance within the energy ranges given in the third column of the table.}
\label{tab:experiments}
\end{center}
\end{table*}

\begin{figure}[t]
\centering
{\large \underline{\textbf{Considering only a xenon experiment}}\par\medskip}
\hspace*{-1.22cm}
\includegraphics[scale=0.97]{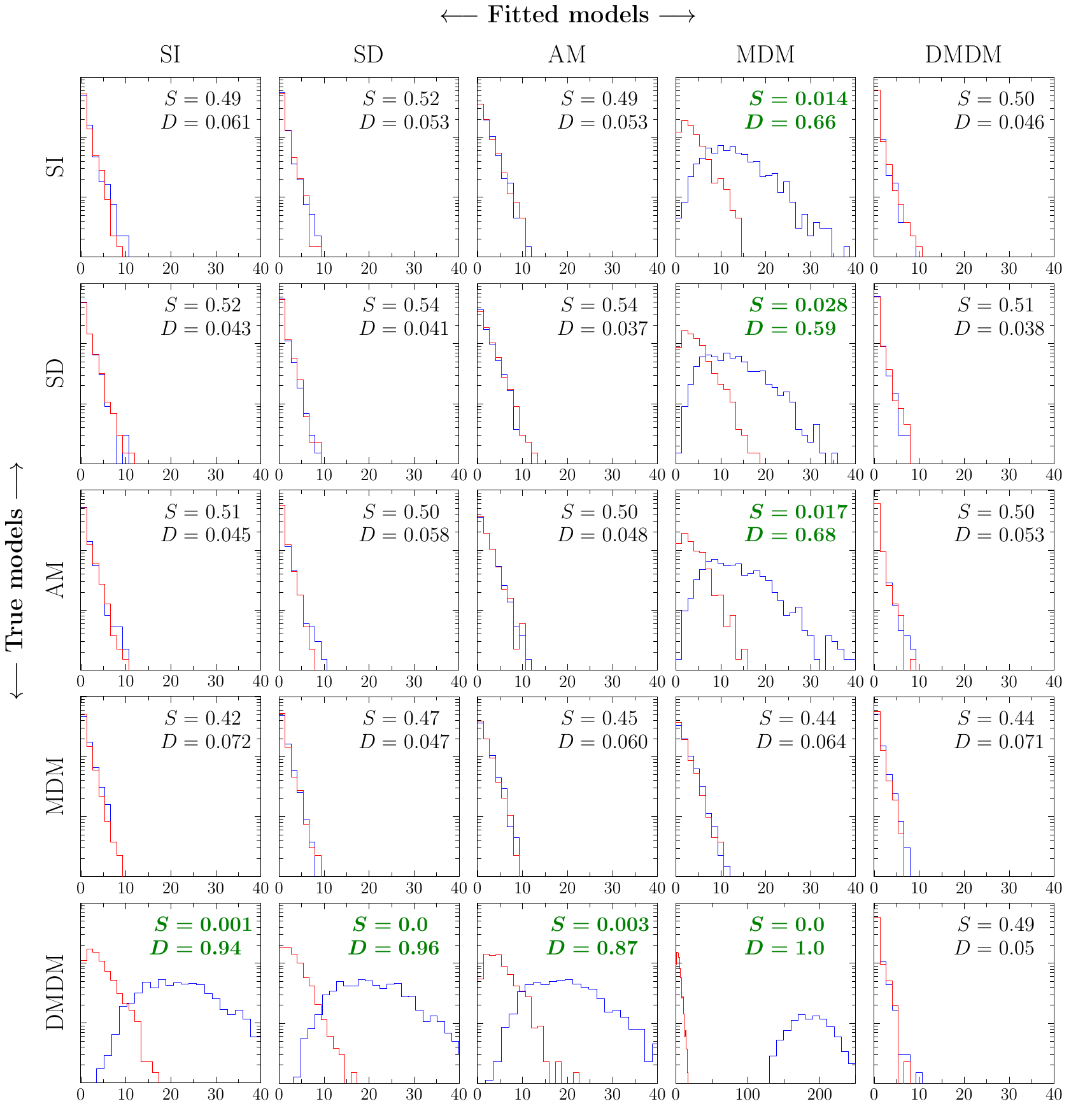}
\caption{\small Distribution of $x_0$ (blue) and $x_1$ (red) for different combinations of true models (rows) and fitted models (columns), using only the information from a future xenon-based experiment. Following section~\ref{sec:models}, we consider spin-independent (SI) and spin-dependent (SD) interactions, as well as DM with an anapole moment (AM), a magnetic dipole moment (MDM), or a dark magnetic dipole moment (DMDM). A pair of models can be distinguished in a halo-independent way if the distributions of $x_0$ and $x_1$ are sufficiently different, leading to a small (large) value of the similarity $S$ (the distinguishability $D$). We indicate in green the scenarios in which $S < 0.05$ or $D > 0.5$, corresponding to an exclusion of the fitted model in a typical realisation of the true model with at least $95\%$ C.L.}
\label{fig:xe_panel}
\end{figure}

The results of the analysis discussed above are shown in figure~\ref{fig:xe_panel}. The different rows correspond to different choices of the \emph{true model}, while the different columns correspond to different \emph{fitted models}. In each case, the blue line shows the distribution of $x_0$, while the red line shows the distribution of $x_1$. We also show the resulting values for $S$ and $D$ in each panel.

\subsection{Discussion}
\label{sec:singleexp_discussion}

We can make a number of interesting observations from figure~\ref{fig:xe_panel}. First of all, we find SI interactions and SD interactions to be completely indistinguishable in our set-up. This result is unsurprising given that both types of interactions have the same momentum and velocity dependence and the differences arising from nuclear form factors can easily be compensated by adjusting the velocity integral accordingly~\cite{Pato:2011de,Peter:2013aha,Kavanagh:2014rya}.

More surprisingly, we find that both SI and SD are also completely indistinguishable from anapole interactions, even though these interactions depend on both the DM velocity and the momentum transfer in a more complicated way. In other words, if DM scattering is due to an anapole moment, astrophysical uncertainties may lead us to misidentify such an interaction as SI or SD scattering (and vice versa).

This result can be understood as follows. Anapole interactions can be split into two separate parts (see section~\ref{sec:models}): one part proportional to the second velocity integral $h(v)$ and one part proportional to the first velocity integral $g(v)$ and the momentum transfer $q^2$. The former contribution predicts a monotonically decreasing spectrum, which differs from SI interactions only because it depends on $h(v)$ rather than $g(v)$, while the latter contribution predicts a spectrum that peaks at finite recoil energy and therefore should lead to strikingly different predictions. It turns out, however, that for a xenon target the relevant nuclear form factor for the second contribution (which is approximately proportional to the magnetic dipole moment of the nucleus) is much smaller than the SI form factor relevant for the first contribution, so that experimental predictions are completely dominated by the first contribution~\cite{Gresham:2014vja}. For a xenon-based experiment anapole and SI interactions therefore only differ in their dependence on the DM velocity and cannot be distinguished without making an assumption on the velocity integrals $g(v)$ and $h(v)$. As we will see below, this conclusion can change for different target materials.

\afterpage{\FloatBarrier}
For magnetic dipole interactions a more non-trivial result is obtained. We find that, if DM scattering is described by SI, SD or anapole interactions, it will be possible to exclude scattering via a magnetic dipole moment at high confidence level. The distinction does however not work in the opposite direction. If DM scattering is due to a magnetic dipole moment, standard SI or SD interactions would still give a good fit to the data (when allowing the DM velocity distribution to vary).

The reason for this observation is that magnetic dipole interactions are enhanced for small recoil energies proportional to inverse powers of the momentum transfer (because they result from the exchange of massless photons). As a result, magnetic dipole interactions fall off more steeply towards larger momentum transfer than SI or SD interactions (for similar velocity integrals). It is therefore possible to interpret scattering from MDM interactions in terms of SI or SD interactions by considering a more steeply falling velocity integral (possibly combined with a smaller value of the DM mass).

\begin{figure}[t]
\centering
\includegraphics[width=0.47\textwidth]{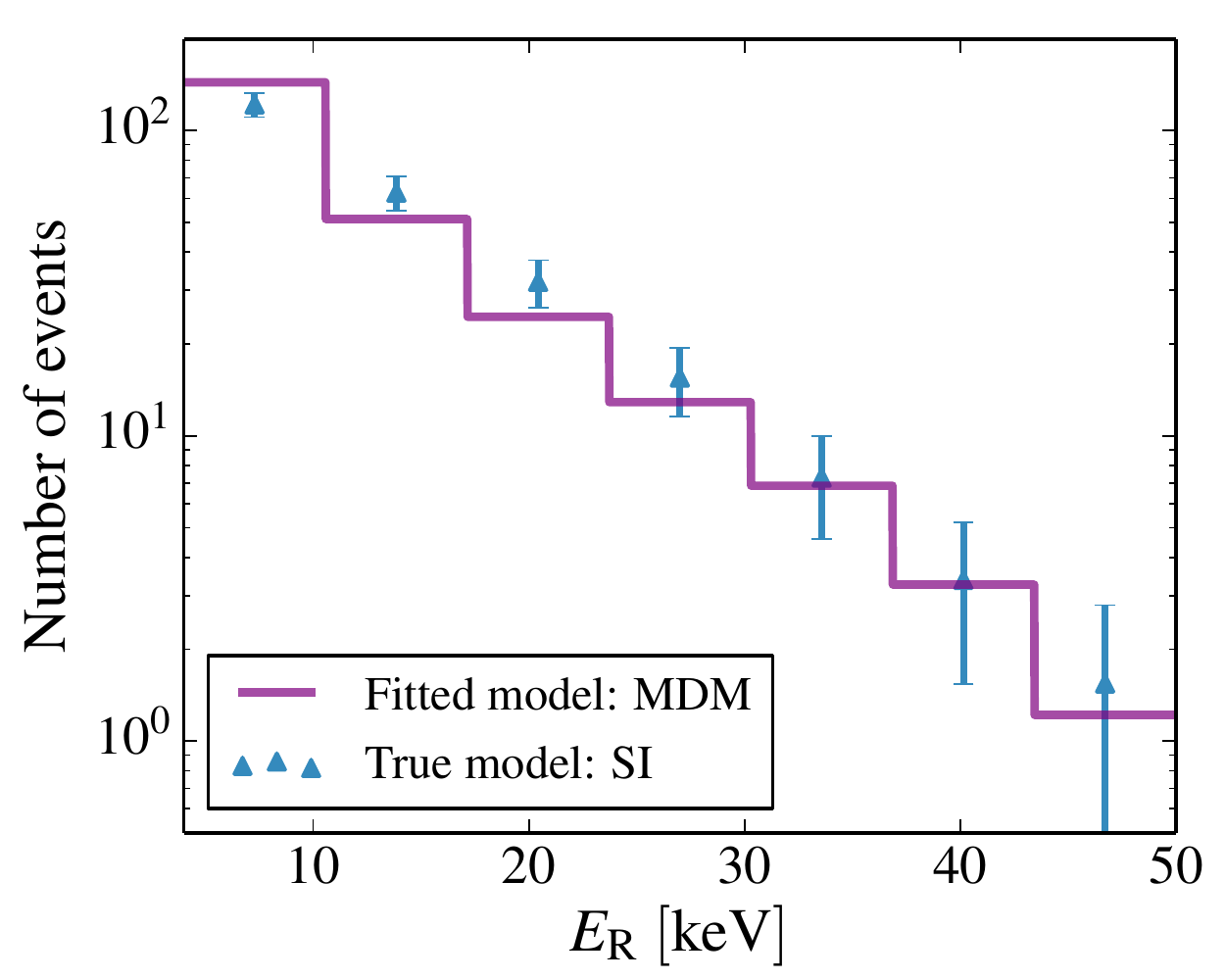}\qquad
\includegraphics[width=0.47\textwidth]{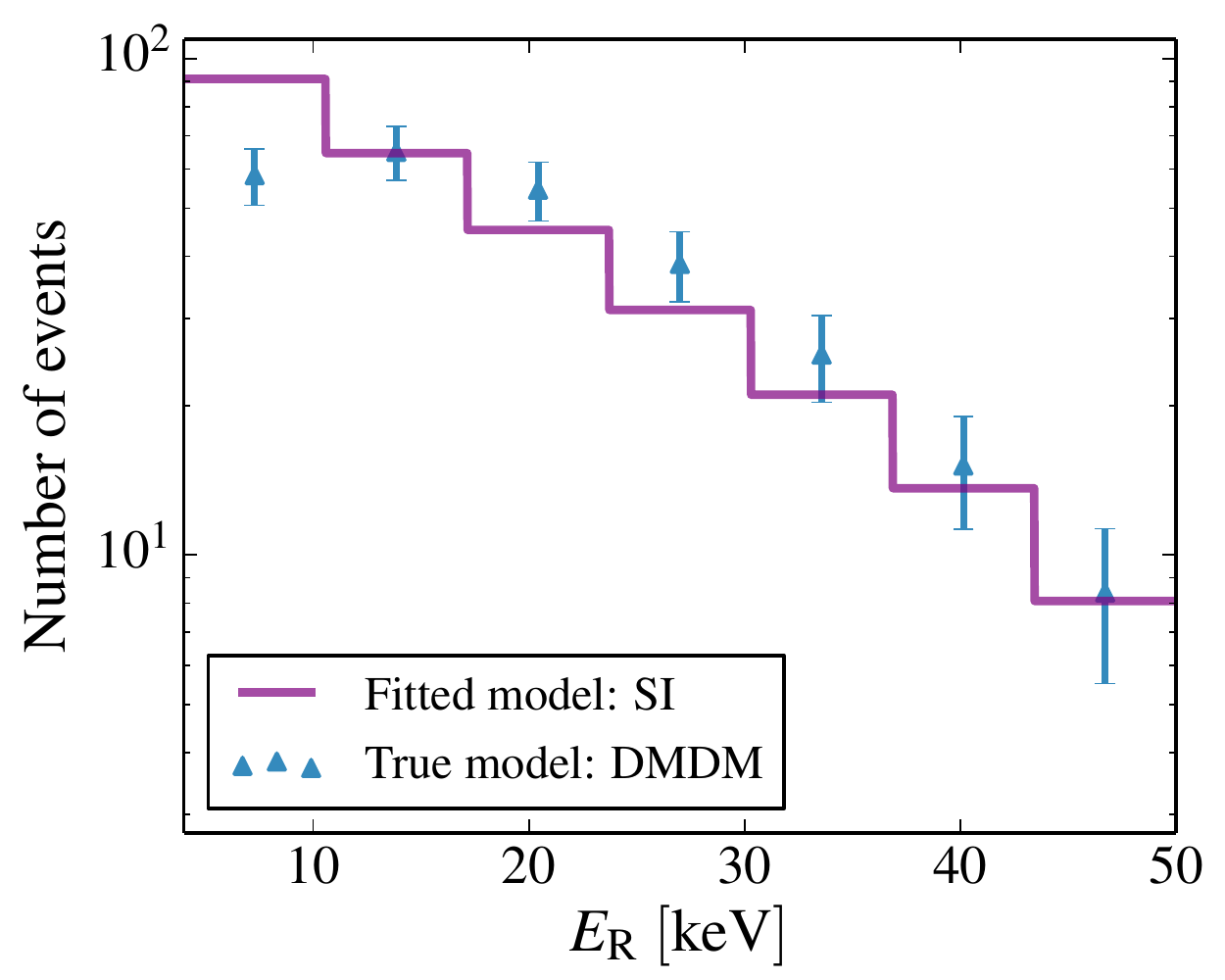}
\caption{Comparison of the predicted event rates and the best-fit event rates for two different combinations of true model and fitted model. For the predictions, we show error bars of size $\sqrt{N}$, where $N$ is the predicted number of events. In these plots no Poisson fluctuations are included, so prediction and best-fit model should agree perfectly if the true model were identical to the fitted model.}
\label{fig:eventrates}
\end{figure}

To fit a recoil spectrum from SI or SD interactions with MDM interactions, on the other hand, would require a flatter velocity integral and a larger value of the DM mass. However, the monotonicity requirement of the velocity integral implies that the recoil spectrum for MDM interactions cannot be arbitrarily flat. In fact, even if the velocity integral is as flat as possible (i.e.\ effectively constant), the predicted recoil spectrum is still too steep to give a good fit to the assumed signal. This expectation is confirmed explicitly in the left panel of figure~\ref{fig:eventrates}, where we show the binned event rates predicted for SI interactions compared to the typical best fit obtained from MDM interactions (as defined above).

For scattering via a dark magnetic dipole moment, we find the converse situation: If DMDM corresponds to the true model of DM, all the other kinds of interaction considered in our study can be ruled out at high confidence level. The reason for this observation is obvious: As shown in the right panel of figure~\ref{fig:eventrates}, DMDM interactions predict a non-monotonic recoil spectrum, i.e.\ a peak at finite recoil energy (see figure~\ref{fig:eventrates}). Given the monotonicity requirement of the velocity integral, such a spectrum can never be fitted with the other interactions that we consider here (see also~\cite{Fox:2010bu}). Our study shows that the non-monotonic nature of the spectrum can be established with a relatively small number of observed events, i.e.\ within the level of sensitivity achievable by upcoming direct detection experiments. In this particular scenario, it would therefore be possible to rule out standard SI or SD interactions at high statistical significance without the need to make any assumptions on the astrophysical distributions of DM.

We emphasise that taking DMDM interactions as the fitted model can nevertheless give a good fit to experimental data resulting from other kinds of interactions. The reason is that it is still possible to obtain a monotonically falling recoil spectrum from DMDM interactions, provided the velocity integral falls very steeply. This explains why no distinguishability is found in the final column of figure~\ref{fig:xe_panel}.

To conclude this discussion we note that, although not shown in figure~\ref{fig:xe_panel}, we have also considered DM interactions due to an electric dipole moment or a dark electric dipole moment (i.e.~a new heavy mediator coupling to a DM electric dipole moment). We have found these cases to be indistinguishable from MDM and DMDM interactions, respectively, and to yield the same conclusions concerning their distinguishability from other kinds of interactions. As these interactions offer no additional insights, we omit them from figure~\ref{fig:xe_panel} and similar figures below.

\subsection{Different couplings to protons and neutrons}
\label{sec:singleexp_different_fpfn}

So far we have only considered very specific coupling structures. As described in section~\ref{sec:models}, we have always assumed isoscalar couplings ($f_p = f_n$) for the SI case, isovector couplings ($g_p = - g_d$) for the SD case and photon-like couplings to the electromagnetic charge for the other cases. Nevertheless, there are many models (such as DM interacting via a $Z'$ with kinetic mixing and mass mixing~\cite{Frandsen:2011cg}), where the interactions of DM can have a more complicated structure. To conclude this section, we therefore relax our assumptions on the coupling structures.

Ref.~\cite{Feldstein:2014gza} has studied the question to what degree the ratio $f_n / f_p$ can be determined from future experiments in a halo-independent way. The conclusion was that, by combining information from several experiments, it may be possible to distinguish between SI interactions mediated by the Higgs ($f_p = f_n$) from SI interactions mediated by the $Z$-boson ($f_p \sim 0$). Here we want to focus on a different question: Could DM interactions via a non-standard DM operator be misidentified as SI interactions with non-isoscalar couplings? In other words, are there situations where standard SI interactions give a bad fit to the data, but an acceptable fit could be obtained for non-isoscalar couplings?

A particularly interesting case to consider in this context is the one where DM has DMDM interactions. As discussed above, in this case it should be possible to rule out standard SI interactions at high significance. Let us now repeat the analysis from above by treating the ratio $f_n / f_p$ as an additional free parameter.

At first sight, changing the ratio of the couplings to protons and neutrons should not significantly alter the shape of the recoil spectrum and should therefore not improve the fit to a non-monotonic recoil spectrum. The crucial observation, however, is that in heavy nuclei like xenon the distribution of protons and neutrons in the nucleus differ slightly, with neutrons having a higher density close to the surface (the so-called neutron skin)~\cite{Zheng:2014nga}. As a result, the form factor for protons falls off more slowly towards larger momentum transfer than the one for neutrons.

The difference between the form factors is completely negligible for $f_p = f_n$, but it plays an important role in the case of destructive interference between the two contributions (sometimes called isospin-violating DM~\cite{Feng:2011vu}). Due to the different form factors, the amount of interference then depends sensitively on the nuclear recoil energy. For example, for \mbox{$f_n / f_p \approx -0.7$}, the cancellation between the two contributions is maximal for zero recoil energy and becomes less important for larger recoil energies, so that the predicted recoil spectrum obtains a non-standard shape~\cite{Zheng:2014nga}. In other words, the form factor differences can potentially lead to a non-monotonic recoil spectrum from SI interactions with $f_n / f_p < 0$.

\begin{figure}[tb]
\centering
\includegraphics[width=0.35\textwidth,clip,trim=0 -9 0 0]{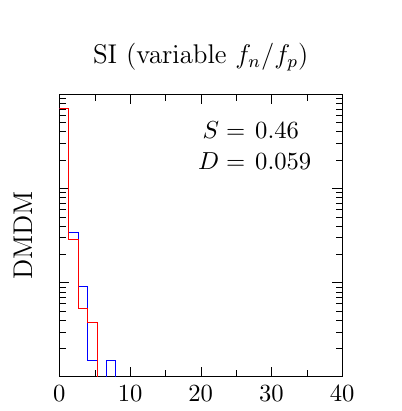}
\qquad
\includegraphics[width=0.47\textwidth]{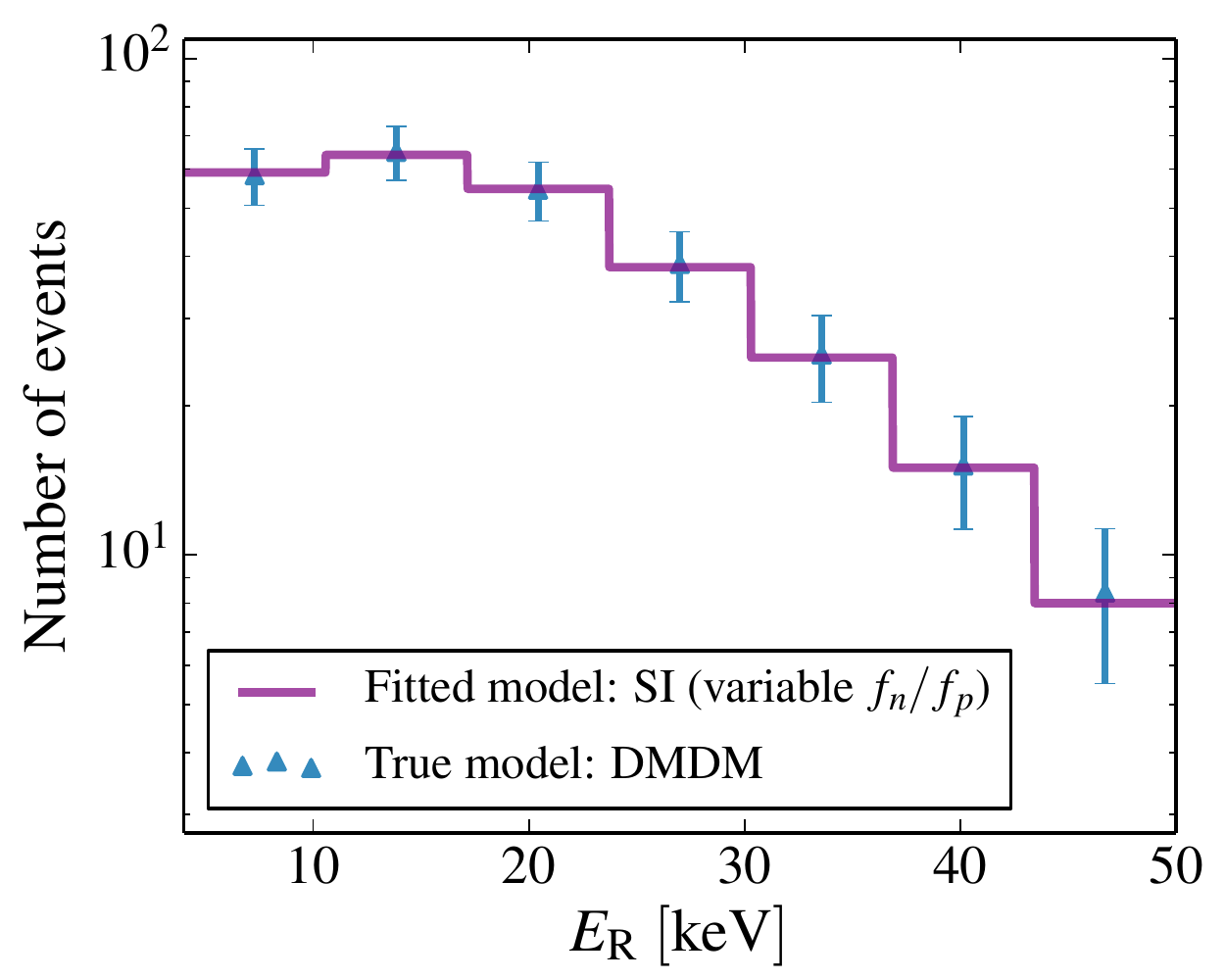}
\caption{Left: Same as figure~\ref{fig:xe_panel} (last row, first column), but allowing varying $f_n / f_p$. Right: Same as figure~\ref{fig:eventrates} (right), but allowing varying $f_n / f_p$.}
\label{fig:fpfn}
\end{figure}

Our results are shown in figure~\ref{fig:fpfn}. We find that, allowing variable $f_n / f_p$, it is indeed possible to fit a non-monotonic recoil spectrum like the one obtained from a dark magnetic dipole moment with SI interactions. As expected, the best-fit point is then found to be close to $f_n / f_p \approx -0.7$. We conclude that these two models are essentially indistinguishable and that DMDM interactions may therefore incorrectly be identified as isospin-violating DM (or vice versa).

It should be clear however that such a confusion can only occur if DM is only observed in a single experiment. If the non-monotonic spectrum does indeed result from an almost maximal destructive interference, much larger event rates would be expected in other target materials with different ratios of protons to neutrons, enabling us to easily test such a scenario. We will quantitatively confirm this statement in section~\ref{sec:severalexp_different_fpfn}.

\section{Constraining DM coupling structures with several experiments}
\label{sec:2exp}

We have seen in the previous section that in many cases a single liquid-xenon based experiment is insufficient to distinguish different coupling scenarios. In this section, we study whether better discrimination can be achieved if a DM signal is seen in more than one kind of experiment. In principle, the presence of several experiments is expected to lead to significant improvements, as in general different target materials have different form factors and therefore both the shape and the normalisation of the spectrum will depend on the coupling structure in different ways for different experiments.\footnote{We note that, furthermore, considering more than one experiment tremendously improves the ability to constrain the DM mass in a halo-independent way, see e.g.~\cite{Peter:2011eu,Kavanagh:2012nr}. Although we do not consider the reconstruction of the DM mass in the present context, we emphasise that this reason alone makes the combination of data from several experiments a very important undertaking.} A trivial example would be an argon-based target, which has no sensitivity to SD interactions and could therefore allow perfect discrimination between SI and SD scattering. 

A more interesting situation can occur for more complicated models like DM with a magnetic dipole moment, where more than one operator contributes in the non-relativistic limit. In this case, the relative contribution of the different operators may vary from target to target, potentially making it possible to distinguish such a scenario from standard interactions. An example for such a case was already discussed in section~\ref{sec:vminspace} (see figure~\ref{fig:vminspace2}). We now use the quantitative methods introduced in section~\ref{sec:gof} to study this and other interesting cases more closely.

In addition to the liquid-xenon-based experiment discussed above, we will consider two further target materials representative of next-generation experiments: a germanium (Ge) semiconductor experiment such as EDELWEISS-III~\cite{Arnaud:2016tpa} or SuperCDMS~\cite{Calkins:2016pnm} and a low-background iodine (I) crystal scintillator such as SABRE~\cite{Shields:2015wka}.\footnote{As discussed above, an argon-based experiment would trivially allow to distinguish between SI and SD interactions. Due to its large threshold however it does not improve the distinguishability for cases with non-standard momentum and velocity dependence and is therefore of limited interest in the present context. Similarly, we also do not consider fluorine experiments, since the spectral information is of crucial information in order to distinguish between different models and it is very difficult for bubble-chamber experiments to infer the nuclear recoil energy.} Our implementation of both types of experiments is summarised in table~\ref{tab:experiments}. As before, we make the simplifying assumption that background contributions can be neglected.

\subsection{Combination of xenon and germanium}

\begin{figure}
\centering
{\large \underline{\textbf{Combination of xenon and germanium}}\par\medskip}
\hspace*{-1.22cm}
\includegraphics[scale=0.98]{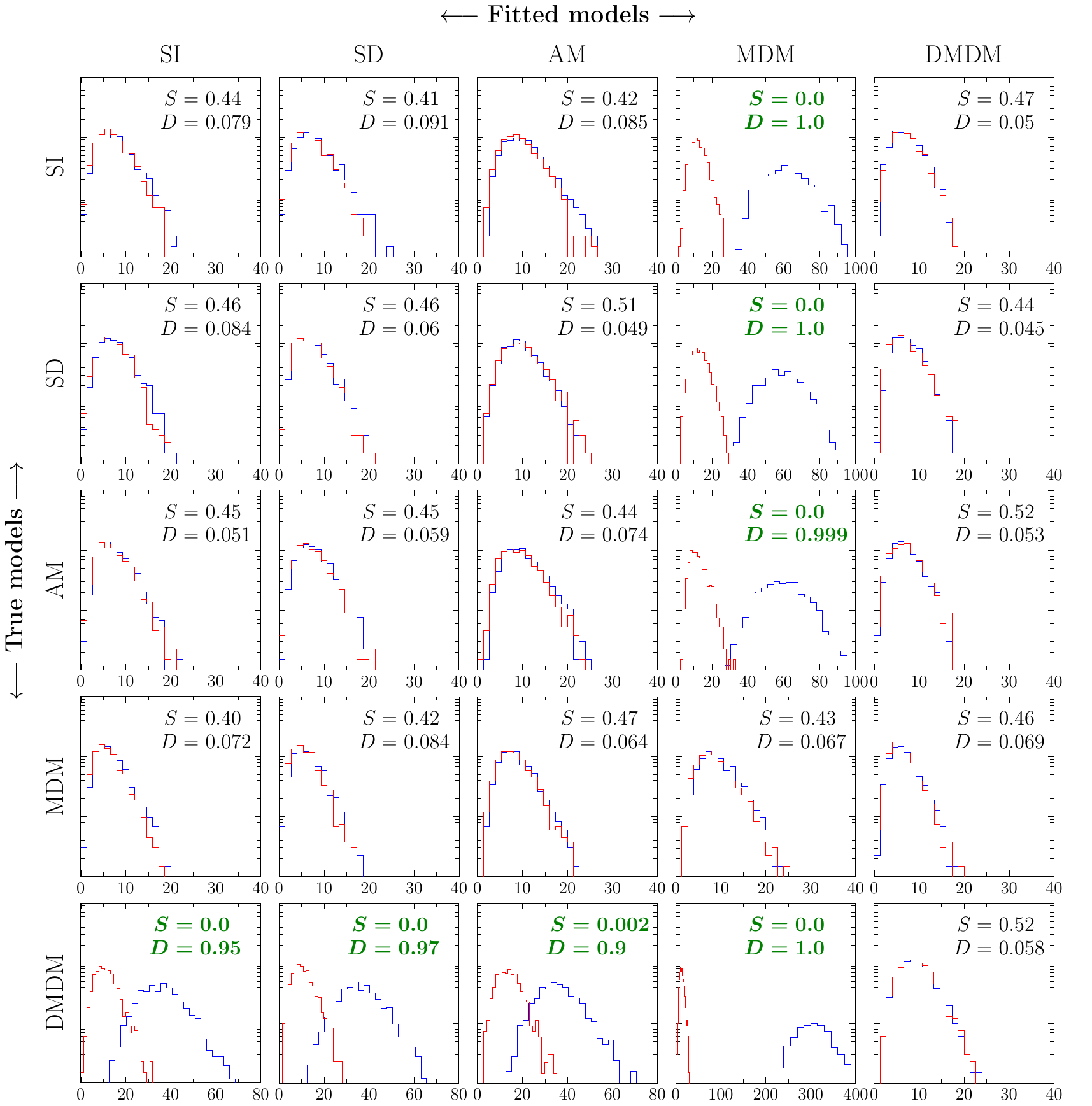}
\caption{Same as figure~\ref{fig:xe_panel}, but considering data from both  a xenon and a germanium target. Again, a combination of a true and fitted model can be distinguished in a halo-independent way if the distributions of $x_0$ (blue) and $x_1$ (red) are sufficiently different, quantified by a small similarity $S$ and/or a large distinguishability $D$.}
\label{fig:xege_panel}
\end{figure}

Let us first discuss to what extent different coupling scenarios can be distinguished in a halo-independent way by combining a xenon- and a germanium-based experiment. To this end we show in figure~\ref{fig:xege_panel} the results of our analysis for the five benchmark models introduced in section~\ref{sec:models}, using the  experimental set-ups given in table~\ref{tab:experiments}. Analogously to figure~\ref{fig:xe_panel}, the blue lines in each panel show the distribution of $x_0$, defined as the minimal $\chi^2$ obtained in the fit of a given fitted model (corresponding to the different columns) to the mock data generated by a particular true model (corresponding to the different rows). Furthermore, the distribution of $x_1$ (shown in red) represents the quality of the fit to the typical prediction of the fitted model (see section~\ref{sec:gof}). 

We can make two immediate observations from figure~\ref{fig:xege_panel}: Firstly, the fourth column shows that the MDM scenario does not provide a good fit to the data generated for any other model. Secondly, the last row implies that if the true model of DM is given by DMDM interactions, one can distinguish it with high confidence level from any of the other scenarios, in particular from the standard SI or SD interactions. Comparing with figure~\ref{fig:xe_panel}, we observe that these pairs of models could already be distinguished with a xenon-based experiment alone (albeit in some cases with a smaller value of the distinguishability parameter $D$ and thus with larger similarity $S$), and we refer to section~\ref{sec:singleexp_discussion} for a physical interpretation of these cases.

Most importantly, we find that adding the germanium target on top of the xenon-based experiment does not allow for the discrimination of any additional pair of models. At first sight, this is unexpected: in general, the individual target nuclei have different relative sensitivities to different models, and hence combining the information from two or more experiments should break any degeneracy between them. For the combination of xenon and germanium targets, however, we find that even increasing the germanium exposure by a factor of 10 (such that the number of observed events in xenon and germanium are comparable) the distinguishability does not improve significantly. Let us therefore pause our discussion of figure~\ref{fig:xege_panel} to discuss a simple argument to make sense of this observation, which is based on the method for halo-independent comparison introduced in~\cite{Fox:2010bz}.

Assuming that for a given DM mass the two experiments based on xenon and germanium are sensitive to a similar range of $v_\text{min}$ values, the rates expected for SI couplings can schematically be written as 
  \begin{align}
  \text{Rate}_{\text{SI,Xe}} \sim (\text{SI sensitivity})_\text{Xe} \, &\times \, g(v_\text{min})  \,,\nonumber \\
  \text{Rate}_{\text{SI,Ge}} \sim (\text{SI sensitivity})_\text{Ge} \, &\times \, g(v_\text{min}) \,.
  \end{align}
Similarly, the rates induced by SD couplings are given by
    \begin{align}
  \text{Rate}_{\text{SD,Xe}} \sim (\text{SD sensitivity})_\text{Xe} \, &\times \, g(v_\text{min}) \,, \nonumber \\
  \text{Rate}_{\text{SD,Ge}} \sim (\text{SD sensitivity})_\text{Ge} \, &\times \, g(v_\text{min}) \,.
  \end{align}
In a region of $v_\text{min}$-space accessible to both experiments, we can define the ratio of rates for SD interactions as
\begin{equation}
 \RR{SD}{Xe}{Ge} \equiv \frac{\text{Rate}_{\text{SD,Xe}}}{\text{Rate}_{\text{SD,Ge}}} = \frac{(\text{SD sensitivity})_\text{Xe}}{(\text{SD sensitivity})_\text{Ge}}
\end{equation}
and analogously $\RR{SI}{Xe}{Ge}$ for SI interactions. 

In order for SI and SD interactions to be distinguishable in a halo-independent way, it is therefore necessary that $\RR{SI}{Xe}{Ge}$ is sufficiently different from $\RR{SD}{Xe}{Ge}$, so that only one of these scenarios can give a good fit to the data for the assumed DM mass.\footnote{Note that even if this is the case, the second scenario may still give a good fit to the data for a different choice of the DM mass, which can change both the ratio of rates and the overlap in $v_\text{min}$-space. In other words, having different ratios of rates is a necessary but not sufficient condition for two interactions to be distinguishable in a halo-independent way.} It turns out, however, that numerically the ratios of the sensitivities are, by coincidence, very similar:
 \begin{align}
  \RR{SI}{Xe}{Ge} &\simeq \frac{A_\text{Xe}^2 \mu_\text{Xe}^2}{A_\text{Ge}^2 \mu_\text{Ge}^2} \approx\ 3.2 \frac{\mu_\text{Xe}^2}{ \mu_\text{Ge}^2} \label{eq:SIratio_XeGe} \\
\RR{SD}{Xe}{Ge} &\simeq \frac{\sum_{T \in \left\{ 129\text{Xe}, 131\text{Xe} \right\}}\eta_{T} (S_p^{(T)} - S_n^{(T)})^2(J_T+1)/J_T }{\eta_{73\text{Ge}} (S_p^{(73\text{Ge})} - S_n^{(73\text{Ge})})^2(J_{73\text{Ge}}+1)/J_{73\text{Ge}} } \approx 2.9 \frac{\mu_\text{Xe}^2}{ \mu_\text{Ge}^2} \,, \label{eq:SDratio_XeGe}
 \end{align}  
where $\eta_T$ is the natural abundance of the target isotope $T$, $S_p^{(T)}$ and  $S_n^{(T)}$ are the expectation values of the spin-content of the proton and neutron group in the nucleus $T$, respectively, and $J_T$ is its total spin. We thus find
\begin{equation}
 \RR{SI}{Xe}{Ge} \approx 1.1 \, \RR{SD}{Xe}{Ge} \; .
 \label{eq:RRratio}
\end{equation}

While this argument is of course simplified, as it does not take into account effects arising e.g.\ from Poisson fluctuations in the data, from the form factor suppression at finite $q^2$, or from the fact that usually one has several bins with different degree of overlap in the $v_\text{min}$ space, equation~(\ref{eq:RRratio}) still gives a qualitative explanation for the similarity of the SI and SD interactions observed in figure~\ref{fig:xege_panel}. 

Returning to the discussion of figure~\ref{fig:xege_panel}, we observe that also in the cases where the true model is given by AM or MDM interactions, a good (halo-independent) fit to the data can be obtained for standard SI or SD interactions. Consequently, one cannot discriminate between these scenarios. The reason is similar to the one discussed in section~\ref{sec:singleexp_discussion} for the xenon-based experiment alone. For scattering off germanium the cross section induced by the anapole moment is dominated by the $v_\perp^2$ term, with a dependence on the target nucleus similar to the SI case. Hence, $\RR{AM}{Xe}{Ge}$ is very similar to the corresponding ratios for SI or SD interactions. The different dependence on the DM velocity does not increase the distinguishability, as they can be fully compensated by choosing an appropriate velocity distribution in the fit. A similar argument holds when replacing the AM interaction by the MDM interaction as the true model. Again, we have confirmed that our conclusions do not change substantially when increasing the exposure of the germanium experiment by a factor of 10, and are hence ``intrinsic'' to this set of target nuclei.

Lastly, we observe that all models under discussion can be well fitted to by the DMDM model, as shown in the final column of figure~\ref{fig:xege_panel}. This observation seems surprising, because for DMDM interactions event rates are proportional to $q^4 = 16 \, v_\text{min}^4 \, \mu_{T\chi}^4$, so that scattering is enhances for heavy targets: $\RR{DMDM}{Xe}{Ge} \gg \RR{SI}{Xe}{Ge}$. One would thus expect the information from the two different targets to be highly complementary. It turns out, however, that in this particular case the difference in the ratio of rates can be fully compensated by changing the assumed DM mass. For example, for true model SI and fitted model DMDM, a good fit to the data from both xenon and germanium can typically be obtained for \mbox{$m_\chi \approx 10\:\text{GeV}$} (compared to an assumed true DM mass of $50\:\text{GeV}$). As for the case of a xenon target alone, the additional momentum dependence of the DMDM interaction is then compensated by a steeply falling velocity integral.

\subsection{Different couplings to protons and neutrons}
\label{sec:severalexp_different_fpfn}

Let us now turn to an example where adding the information from a germanium target makes an important difference for the discrimination of different models. For this purpose we return to the question of whether a non-standard DM-nucleon interaction such as DMDM could be confused with an SI scenario with variable neutron-to-proton coupling ratio $f_n/f_p$. In section~\ref{sec:singleexp_different_fpfn} we found that due to the slightly different form factors for protons and neutrons, the recoil spectrum in a single experiment can be non-monotonic even for SI interactions, provided that $f_n/f_p$ is close to the value leading to maximal destructive interference. In particular we concluded from figure~\ref{fig:fpfn} that an SI interaction with variable neutron-to-proton coupling ratio can provide a good (halo-independent) fit to mock data generated for DMDM interactions (considering only a xenon experiment).

\begin{figure}[tb]
\centering
\includegraphics[width=0.35\textwidth,clip,trim=0 -10 0 0]{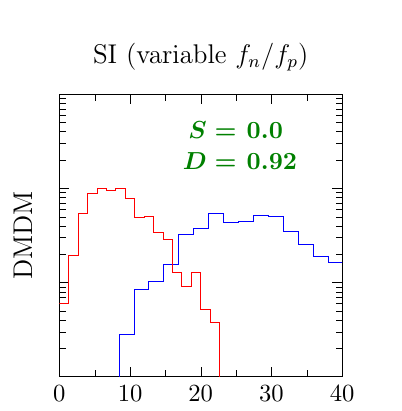}
\qquad
\includegraphics[width=0.47\textwidth]{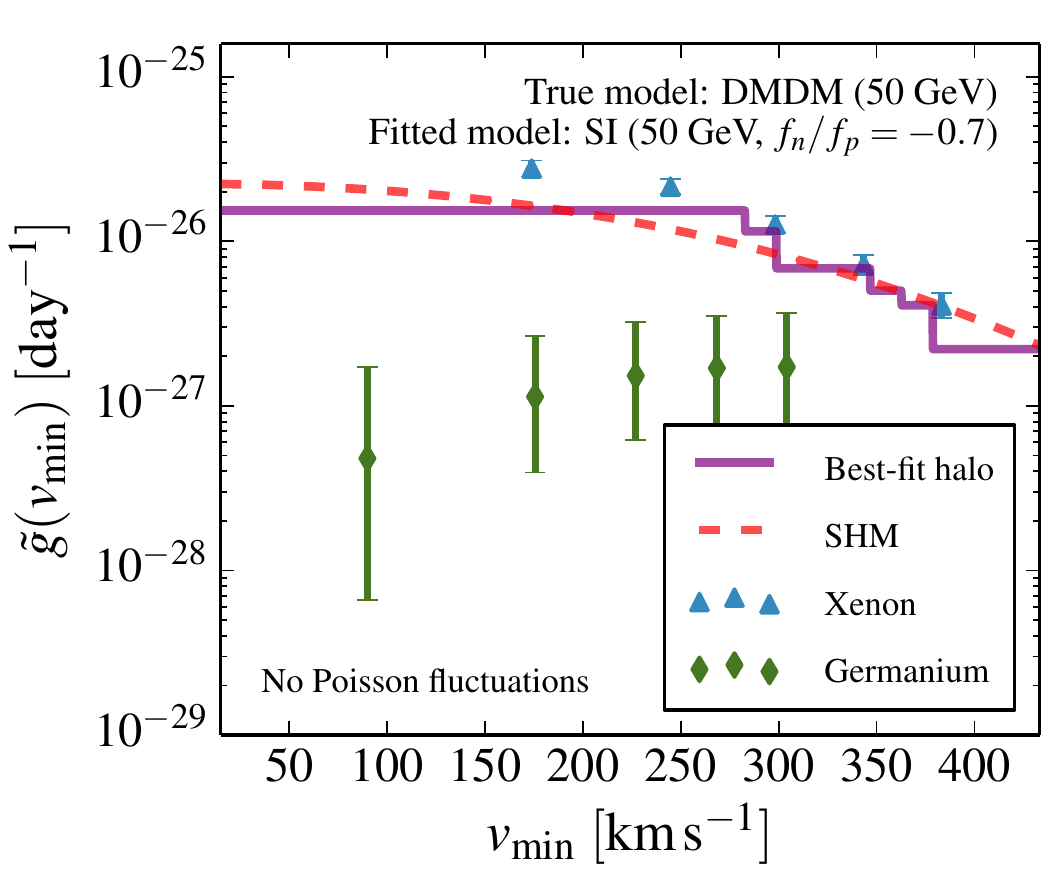}
\caption{Left: The distributions of $x_0$ and $x_1$ for the case that the true model corresponds to DMDM interactions and the fitted model is SI with variable coupling ratio $f_n/f_p$, using the combined information from a xenon and a germanium experiment. Right: Illustration of the best-fit velocity integral obtained for $f_n / f_p = -0.7$. The mapping to $v_\text{min}$-space is described in section~\ref{sec:vminspace} (see also figures~\ref{fig:vminspace1} and~\ref{fig:vminspace2}).}
\label{fig:fpfn_XeGe}
\end{figure}

The value of $f_n/f_p$ leading to maximally destructive interference does however differ from one target nucleus to the other: for example, it is approximately $-0.7$ for xenon, but $-0.79$ for germanium. In particular, we find that there is no value of the neutron-to-proton coupling ratio for which the recoil spectrum expected from SI interactions is non-monotonic \emph{both} at a xenon- and a germanium-based experiment. Consequently, we expect the combination of these two experiments to be able to rule out an SI interaction scenario with variable $f_n/f_p$, when the true model is DMDM. This expectation is confirmed in figure~\ref{fig:fpfn_XeGe}, where we show the corresponding distributions of $x_0$ and $x_1$. Clearly there is a large halo-independent distinguishability of the two interaction scenarios.

An intuitive way to understand this result is presented in the right panel of figure~\ref{fig:fpfn_XeGe} using the mapping to $v_\text{min}$-space introduced in section~\ref{sec:vminspace}. For this figure we have generated mock data (without Poisson fluctuations) based on the DMDM model and then performed the mapping to $v_\text{min}$-space for SI interactions with $f_n / f_p = -0.7$. For the xenon-based experiment, this reconstruction yields monotonically decreasing measurements of the velocity integral, allowing for a consistent fit of all data points. The values of the velocity integral inferred from germanium, however, can be seen to increase towards larger values of $v_\text{min}$ (although for the assumed exposures this increase is not statistically significant). More importantly, since the destructive interference is maximal in xenon, but less pronounced in germanium, the velocity integrals inferred from the two experiments are in strong tension with each other. This tension ultimately allows to exclude the assumed value of $f_n / f_p$ used for the reconstruction in a halo-independent way.

\subsection{Combination of xenon and iodine}

\begin{figure}
\centering
{\large \underline{\textbf{Combination of xenon and iodine}}\par\medskip}
\hspace*{-1.22cm}
\includegraphics[scale=0.98]{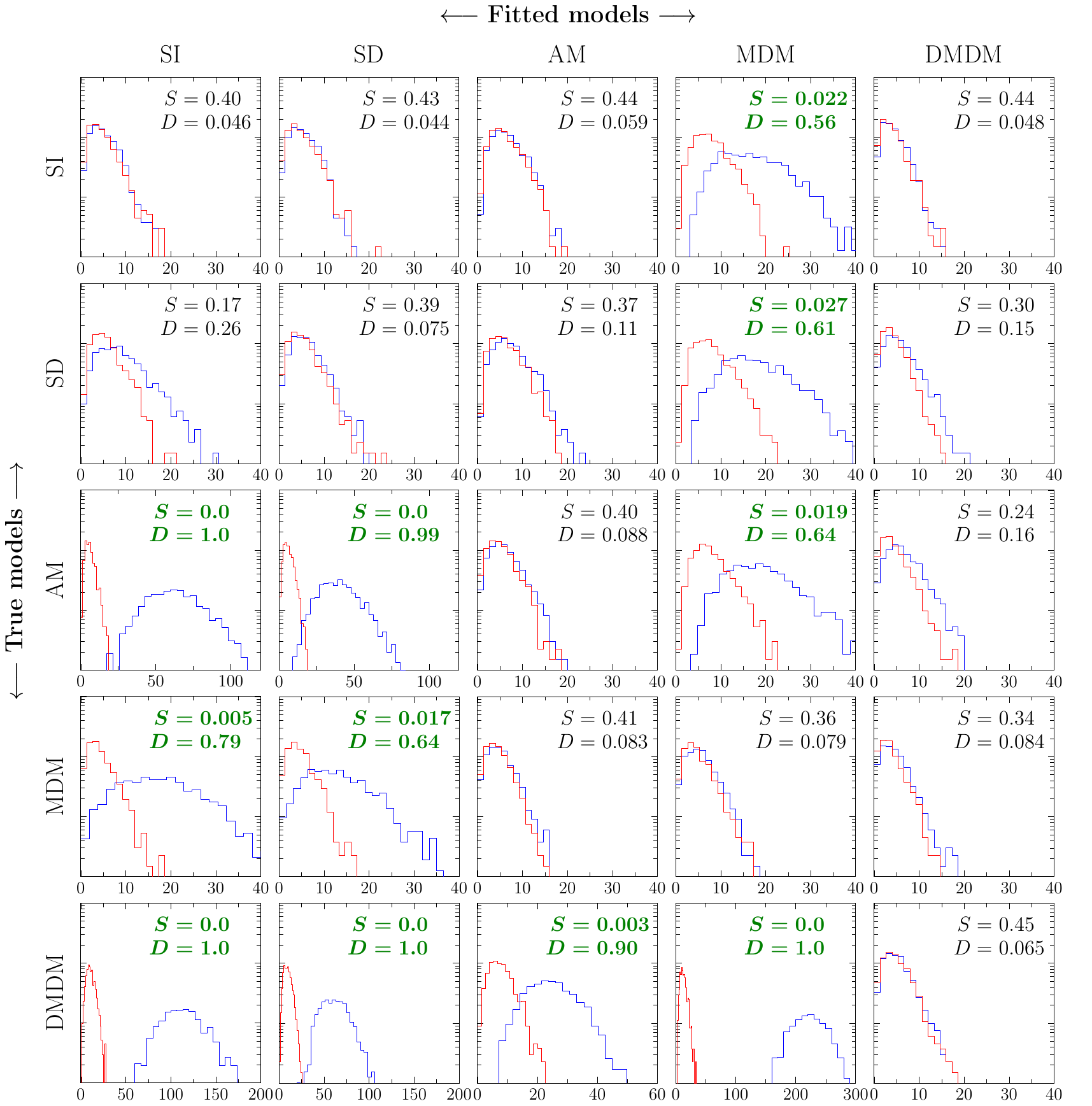}
\caption{Same as figures~\ref{fig:xe_panel} and~\ref{fig:xege_panel}, but considering data from both a xenon and a iodine target.}
\label{fig:xei_panel}
\end{figure}

Let us now turn to the discussion of the results obtained from combining the information of a xenon- and an iodine-based experiment (figure~\ref{fig:xei_panel}). The upper left $2 \times 2$ panels indicate that this set of experiments also cannot distinguish between standard SI and SD interactions with large significance in a halo-independent way. As above, this can be qualitatively explained by estimating the ratios of rates
 \begin{align}
 \RR{SI}{Xe}{I} \approx 1.1 \frac{\mu_\text{Xe}^2}{ \mu_\text{I}^2} \quad \text{and} \quad
 \RR{SD}{Xe}{I} \approx 1.2 \frac{\mu_\text{Xe}^2}{ \mu_\text{I}^2} \,,
 \end{align}
 so that
 \begin{equation}
  \RR{SI}{Xe}{I} \approx 0.9 \, \RR{SD}{Xe}{I} \; .
 \end{equation}

Interestingly, we find that if the true model is given by AM interactions, adding the information from the iodine experiment helps substantially in discriminating this scenario from the standard SI or SD interaction. This observation can be explained by the fact that for AM scattering off iodine the term $\propto q^2$ in the scattering cross section dominates due to the large magnetic dipole moment of the iodine nucleus~\cite{Chang:2010en}. This term gives rise to a peak in the recoil spectrum (see also figure 9 in~\cite{Gresham:2014vja}), which cannot be compensated by adjusting the velocity distribution, leading to a worse fit for models without such a peak. Furthermore, as the $q^2$ contribution is only important for iodine, the ratio of the rates for scattering off xenon and iodine is significantly different for AM and for SI (or SD) interactions, leading to a (halo-independent) tension between the experiments. Together these two observations explain the large distinguishability for the case that the true model has AM interactions and the fitted model is SI or SD.

If, on the other hand, the true model is given by MDM interactions the additional $1/q^2$ term in the cross section induced by the photon propagator always leads to a monotonically decreasing spectrum. Nevertheless, the second argument from above remains valid. The different magnitude of the nuclear magnetic dipole moment for iodine and xenon still leads to a halo-independent tension when incorrectly trying to fit MDM interactions with SI or SD interactions (see figure~\ref{fig:vminspace2}). Hence, there is still some discrimination power ($D=0.79$ and 0.64, respectively) for this case.

\section{Conclusions}
\label{sec:conclusions}

Near-future experiments for the direct detection of DM promise significant improvements in sensitivity over existing searches. In anticipation of these improvements it is timely to develop and refine strategies for extracting the particle physics properties of DM from data. Because of the strong dependence of experimental event rates on the essentially unknown Galactic velocity distribution of DM it is of particular importance to devise \emph{halo-independent} methods, i.e.\ methods that enable us to deduce properties of the DM particle from a positive detection without the need to make any assumptions on the astrophysical distributions.

In this work we have presented a very general framework for analysing the data from one or several direct detection experiments independent of astrophysical assumptions and for quantifying the goodness-of-fit of a given particle physics scenario. Following~\cite{Feldstein:2014gza,Feldstein:2014ufa}, we perform a halo-independent fit to a given set of data by parametrizing the velocity integral as a piecewise constant function with a sufficiently large number of steps. As we determine the best-fit velocity distribution directly from data instead of fixing it to e.g.~a Maxwell-Boltzmann distribution, our approach can be employed in order to test whether there exists \emph{any} velocity distribution for which a given particle physics scenario of DM is compatible with the data. Importantly, we have shown that this formalism can be applied universally to any combination of non-relativistic operators describing the interactions of DM with nuclei, in particular to models with non-standard dependence of the scattering cross section on the momentum exchange and the DM velocity. This enables us to consider a much broader set of particle physics models compared to many existing halo-independent analyses in the literature.

To demonstrate how this method can be applied to the case of a positive detection of DM in one or several experiments, we have studied mock data generated for realistic near-future direct detection experiments. To this end we have considered a range of different possibilities for the true particle physics nature of DM, including the standard spin-independent and spin-dependent interaction as well as scenarios involving a non-trivial velocity and momentum dependence of the DM-nucleon scattering cross section. For each mock data set we then perform halo-independent fits to the data for various different assumptions on the properties of the DM particle (see figures~\ref{fig:vminspace1} and~\ref{fig:vminspace2}). 

By repeating this fitting procedure for a large number of Poisson realisations of mock data sets, we consistently take into account the statistical noise expected in the observed data. We can then determine which of the interaction scenarios can be distinguished without specifying the velocity distribution. To quantify the similarity and distinguishability of different DM interaction scenarios we have developed a simplified procedure based on the \emph{typical predictions} of the fitted model (figure~\ref{fig:illustration}). We have confirmed the validity of this approximation using explicit Monte Carlo simulations (figure~\ref{fig:pvalues}).

Interestingly we find that already a single experiment like XENON1T with a moderate exposure of two ton-years can be employed for inferring non-trivial information about the particle physics properties of DM in a fully halo-independent way (figures~\ref{fig:xe_panel} and~\ref{fig:eventrates}). For example, our results show that a model of DM predicting a non-monotonic recoil spectrum (realized e.g.~for a dark magnetic dipole moment) can be clearly distinguished from the standard spin-independent or spin-dependent interaction scenario with a XENON1T-like experiment. Also, if DM interacts via the standard spin-independent or spin-dependent coupling, it is possible to rule out in a halo-independent way long-range interactions of DM with nucleons, which could be induced e.g.~by a magnetic dipole moment of DM.

Lastly we have studied to what extent a detection of DM in more than one experiment helps in discriminating different particle physics scenarios. We have found that adding a germanium target to the xenon-based experiment does not increase significantly the halo-independent distinguishability of the scenarios discussed in this work (figure~\ref{fig:xege_panel}). We have illuminated the numerical results by a simplified analytical argument explaining why the complementarity of a xenon and germanium target is strongly limited. An exception to this statement is the case in which the data of both experiments are fitted assuming a spin-independent interaction with variable neutron-to-proton coupling ratio $f_n/f_p$, which can be constrained much more tightly by two experiments with different target nuclei (figures~\ref{fig:fpfn} and~\ref{fig:fpfn_XeGe}). On the other hand, we have shown that the presence of a DM signal in an iodine-based experiment in addition to the xenon-based detector yields additional discrimination power (figure~\ref{fig:xei_panel}). In particular, if scattering is induced by an anapole or magnetic dipole moment of DM, it should be possible to rule out the standard assumption of spin-independent or spin-dependent interactions by combining the information from both experiments, again without referring to a particular velocity distribution of DM in the galaxy.

Our analysis is based on rather modest assumptions on the exposures of upcoming experiments. Clearly, stronger discrimination power can be achieved by more ambitious experiments and by combining the information from more than two experiments. Nevertheless, hundred DM events in a liquid xenon experiment and ten events in an additional experiment may already be sufficient to extract highly non-trivial information on the particle physics nature of DM in a completely halo-independent way. While the conclusion may simply be that some more exotic models of DM can be ruled out, future experiments may also point in the opposite direction and tell us that the interactions of DM are much more complicated than what is usually assumed.

\acknowledgments

We thank Julien Billard, David Cerdeno, Brian Feldstein, Mattia Fornasa, Alejandro Ibarra, Miguel Peiro and Andreas Rappelt for discussions. This work is supported by the German Science Foundation (DFG) under the Collaborative Research Center~(SFB) 676 Particles, Strings and the Early Universe, as well as by the Studienstiftung des Deutschen Volkes and the TUM Graduate School.

\appendix

\section{Non-relativistic effective interactions}
\label{ap:non-relativistic}

Here and below, we follow largely the conventions and notation from~\cite{Catena:2015uua}. It has been shown in~\cite{Fitzpatrick:2012ix} that in the non-relativistic limit the interactions between the DM particle and quarks can be written as a linear combination of 15 effective operators
\begin{align}
\mathcal{L}_\text{eff} = \sum_{\lambda = 1}^{15} \sum_{\tau=0}^1 c_\lambda^{(\tau)} \mathcal{O}_\lambda^{(\tau)} \; ,
\label{eq:Leff}
\end{align}
where $\tau = 0$ for isoscalar interactions (i.e.\ equal couplings to protons and neutrons) and $\tau = 1$ for isovector interactions (opposite couplings to protons and neutrons).\footnote{The operator $\mathcal{O}_2$ is typically discarded, because it does not arise at leading order from a relativistic UV completion.} 

Additional operators can be constructed by multiplying this basic set with (inverse) powers of $q^2$. Such composite operators can arise for example from long-range interactions or from derivative interactions and are typically present if DM interacts via an electric or magnetic dipole moment. We include the three simplest cases in our discussion by expanding the coefficients $c_\lambda^{(\tau)}$ from equation~(\ref{eq:Leff}) as
\begin{align}
c_\lambda^{(\tau)} \equiv \frac{m_N^2}{q^2} \, z_{0,\lambda}^{(\tau)} + z_{1,\lambda}^{(\tau)} + \frac{q^2}{m_N^2} \, z_{2,\lambda}^{(\tau)} \,,
\end{align}
where $m_N \equiv m_p \approx m_n$ is the nucleon mass. The interactions of DM with nucleons can therefore be fully described by the $14 \times 2 \times 3$ coefficients $z_{\kappa,\lambda}^{(\tau)}$, which we combine into a coefficient vector $\mathbf{z}$ for simplicity. We note that our formalism assumes that the mediator(s) of DM scattering are either very light or very heavy compared to the typical momentum transfer. We do not discuss the case of a mediator with $m_\text{med}^2 \sim q^2$.

In order to calculate the differential scattering cross section $\td \sigma/\td E_\mathrm{R}$ for a given target nucleus, the next step is to calculate the so-called DM response functions $S_\alpha^{(\tau,\tau')}$ for \mbox{$\alpha = 1,\ldots,8$}, which depend on the coefficient vector $\mathbf{z}$ in a known way. The DM response functions then need to be multiplied with the corresponding nuclear form factors $\widetilde{W}_\alpha^{(\tau,\tau')}$:
\begin{equation}
 \frac{\td \sigma}{\td E_\text{R}} = \frac{m_T}{2 \pi v^2} \sum_{\alpha=1}^{8} \sum_{\tau=0}^1 \sum_{\tau'=0}^1\, S_\alpha^{(\tau,\tau')} \left( \textbf{z}, v^2, q^2 \right) \, \, \widetilde{W}_\alpha^{(\tau,\tau')}(q^2) \,,
\label{eq:crosssec_Ptot}
\end{equation}
where the sum over $ S_\alpha^{(\tau,\tau')} \,  W_\alpha^{(\tau,\tau')}$ is also referred to as the transition probability. We will now discuss the two ingredients for calculating this quantity in detail.

\subsection{Dark matter response functions \texorpdfstring{$\bm{S_\alpha^{(\tau,\tau')}}$}{}}

The DM response functions $S_\alpha^{(\tau,\tau')}$ are defined in appendix A of~\cite{Catena:2015uua}. The index $\alpha = 1,\dots,8$ runs over the eight different response functions in the order as they are defined in equation~(A.1) of~\cite{Catena:2015uua}. The response functions depend on the coefficient vector $\textbf{z}$, the relative velocity $v$ and momentum transfer $q$. The crucial observation for our purposes is that it is possible to decompose the response functions as a sum of a velocity-independent term and a term proportional to $v^2$:
\begin{align}
S_\alpha^{(\tau,\tau')} \left( \textbf{z}, v^2, q^2 \right) \equiv S_{1,\alpha}^{(\tau,\tau')} \left(\textbf{z}, q^2 \right) + v^2 \cdot S_{2,\alpha}^{(\tau,\tau')} \left(\textbf{z}, q^2 \right) \,.
\label{eq:def_S}
\end{align}
The functions $S_{\beta,\alpha}^{(\tau,\tau')} \left(q^2 \right)$ (with $\beta \in \left\{1,2\right\}$) can furthermore be written as a linear combination of integer powers of $q^2$ (see~\cite{Catena:2015uua}):
\begin{align}
S_{\beta,\alpha}^{(\tau,\tau')} \left(\textbf{z}, q^2 \right) = \sum_{l=-2}^5 \sigma_{\beta,\alpha,l}^{(\tau,\tau')} \left(\textbf{z}\right) \cdot a^l \,,
\label{eq:decomp_S}
\end{align}
where we have introduced the dimensionless quantity
\begin{align}
a \equiv \frac{q^2}{m_N^2}
\label{eq:a}
\end{align}
and implicitly defined the coefficients $\sigma_{\beta,\alpha,l}^{(\tau,\tau')}$ for $l = -2,\ldots,5$. These $2 \times 8 \times 8 \times 2 \times 2$ coefficients fully characterise the DM response and can be calculated in terms of the coefficient vector $\mathbf{z}$. This calculation, although cumbersome, does not pose any fundamental difficulties and will therefore not be discussed further. For the remainder of the discussion it will be sufficient to simply think of $\sigma_{\beta,\alpha,l}^{(\tau,\tau')}(\mathbf{z})$ as the quantity describing the particle physics properties of DM. 

\subsection{Nuclear form factors \texorpdfstring{$\bm{\widetilde{W}_\alpha^{(\tau,\tau')}}$}{}}

Apart from the DM response functions, the differential cross section defined in equation~(\ref{eq:crosssec_Ptot}) also depends on the nuclear form factors $\widetilde{W}_\alpha^{(\tau,\tau')}$, which are provided in~\cite{Anand:2013yka}.\footnote{For convenience we define $\widetilde{W}_\alpha^{(\tau,\tau')} \equiv \frac{4\pi}{2 J_A +1} W_\alpha^{(\tau,\tau')}$, where $W_\alpha^{(\tau,\tau')}$ are the functions provided in~\cite{Anand:2013yka}.} These form factors depend on $q^2$ via the combination $y = \frac{1}{4} b_A^2 q^2$, with
\begin{equation}
b_A = 1\:\text{fm} \times \sqrt{41.467/\left(45 A^{-1/3}-25 A^{-2/3}\right)}
\end{equation}
The nuclear form factors are given as fit functions of the form
\begin{align}
\widetilde{W}_\alpha^{(\tau,\tau')}(y) = \frac{4\pi}{2 J_A +1} \, \exp(-2y) \, \sum_{k=0}^{10} \overline{w}_{\alpha,k}^{(\tau,\tau')} \cdot y^k \,.
\end{align}
Defining $y(a) = \frac{1}{4} b_A^2 m_N^2 \cdot a$, this can be rewritten in the form
\begin{align}
\widetilde{W}_\alpha^{(\tau,\tau')}(a) =  \exp(-2y(a)) \, \sum_{k=0}^{10} w_{\alpha,k}^{(\tau,\tau')} \cdot a^k \,,
\label{eq:decomp_W}
\end{align}
with $w_{\alpha,k}^{(\tau,\tau')} = \frac{4\pi}{2 J_A +1} \left( \frac{1}{4} b_A^2 m_N^2 \right)^k \overline{w}_{\alpha,k}^{(\tau,\tau')}$.

\subsection{Decomposing the differential cross section}

Substituting equations~(\ref{eq:def_S}), (\ref{eq:decomp_S}), and~(\ref{eq:decomp_W}) into equation~(\ref{eq:crosssec_Ptot}), it becomes clear that $\td \sigma/\td E_\text{R}$ can be decomposed into
\begin{equation}
 \frac{\td \sigma}{\td E_\text{R}} = \frac{\td \sigma_1}{\td E_\text{R}}\frac{1}{v^2} + \frac{\td \sigma_2}{\td E_\text{R}}
\end{equation}
with
\begin{align}
 \frac{\td \sigma_\beta}{\td E_\text{R}} & = \frac{m_T}{2 \pi} \sum_{\alpha=1}^{8} \, \sum_{\tau=0}^1 \, \sum_{\tau'=0}^1\, S_{\beta,\alpha}^{(\tau,\tau')} \left(\textbf{z}, q^2 \right) e^{-2y(a)} \sum_{k=0}^{10} w_{\alpha,k}^{(\tau,\tau')} \cdot a^k \nonumber \\
& = \frac{m_T}{2 \pi} \sum_{\alpha=1}^{8} \, \sum_{\tau=0}^1 \, \sum_{\tau'=0}^1 \, \sum_{k=0}^{10} \, \sum_{l=-2}^5 \sigma_{\beta,\alpha,l}^{(\tau,\tau')} \left(\textbf{z}\right) \cdot  w_{\alpha,k}^{(\tau,\tau')} \cdot e^{-2y(a)} a^{k+l} \; .
\label{eq:decomp_cs}
\end{align}
The decomposition of the differential cross section into different powers of $v^2$ is essential for the halo-independent formalism discussed in this paper. The additional decomposition into different powers of $a$ will be extremely convenient for reducing the number of numerical integrations necessary for a concrete implementation of this formalism, as discussed in below.

\subsection{Efficient numerical implementation}
\label{ap:splitting}

In order to determine the matrix $D_{ij}$ for a given experiment, we need to calculate the matrices
\begin{align}
G_{ij} & = \frac{\kappa \, \rho}{2 \, m_T \, m_\chi} \int_{E_j}^{E_{j+1}} \frac{\text{d} \sigma_1}{\mathrm{d} E_{\text{R}}} \, \epsilon(E_\mathrm{R}) \left[\text{erf}\left(\frac{E_{i+1} - E_\text{R}}{\sqrt{2} \Delta E_\text{R}}\right)-\text{erf}\left(\frac{E_i - E_\text{R}}{\sqrt{2} \Delta E_\text{R}}\right)\right] \, \mathrm{d}E_\mathrm{R} \; ,
\\
H_{ij} & = \frac{\kappa \, \rho}{2 \, m_T \, m_\chi} \int_{E_j}^{E_{j+1}} \frac{\text{d} \sigma_2}{\mathrm{d} E_{\text{R}}} \, \epsilon(E_\mathrm{R}) \left[\text{erf}\left(\frac{E_{i+1} - E_\text{R}}{\sqrt{2} \Delta E_\text{R}}\right)-\text{erf}\left(\frac{E_i - E_\text{R}}{\sqrt{2} \Delta E_\text{R}}\right)\right] \, \mathrm{d}E_\mathrm{R} \; .
\end{align}
In principle, the necessary numerical integrations are straight-forward once the details of the DM interactions have been specified. Nevertheless, these calculation can be computationally rather expensive, in particular if $N_\mathrm{s}$ is very large. Fortunately it is possible to significantly reduce the number of necessary integrations by splitting the calculation into several separate steps.

Comparing these expressions for $G_{ij}$ and $H_{ij}$ to equation~(\ref{eq:decomp_cs}) and noting that the coefficients $\sigma_{\beta,\alpha,l}^{(\tau,\tau')} \left(\textbf{z}\right)$ and $w_{\alpha,k}^{(\tau,\tau')}$ do not depend on the recoil energy, it becomes clear that we first need to calculate the functions
\begin{align}
d_{i}^{(p)}(a) &\equiv \frac{1}{2} \int_{0}^{E_\text{R}(a)} \td E_\mathrm{R} \, e^{-2\,y(a(E_\mathrm{R}))} \, a(E_\mathrm{R})^p \, \epsilon(E_\mathrm{R}) \left[ \text{erf} \left( \frac{E'_{i+1}-E_\mathrm{R}}{\sqrt{2} \Delta E_\text{R}} \right) - \text{erf} \left( \frac{E'_{i}-E_\mathrm{R}}{\sqrt{2} \Delta E_\text{R}} \right) \right] \nonumber\\[0.15cm]
&= \frac{m_N^2}{4 \, m_T} \int_{0}^{a} \td a' \, e^{-2y(a')} \, a'^p \, \epsilon \Bigl( \frac{a' \,m_N^2}{2 \, m_T} \Bigr) \Biggl[ \text{erf} \biggl( \frac{E'_{i+1}-\frac{a' \,m_N^2}{2 \, m_T}}{\sqrt{2} \Delta E_\text{R}} \biggr) - \text{erf} \biggl( \frac{E'_{i}-\frac{a' \,m_N^2}{2 \, m_T}}{\sqrt{2} \Delta E_\text{R}} \biggr) \Biggr] \,,
\end{align}
where $a = q^2 / m_N^2 = 2 \, m_T \, E_\text{R} / m_N^2$.

These functions can be thought of as the predicted number of events in the $i$th bin for a hypothetical scattering cross section with simplified energy dependence of the form $\mathrm{d}\sigma/\mathrm{d}E_\mathrm{R} \sim \exp(-E_\mathrm{R}) E_\mathrm{R}^p$ and a velocity integral of the form $g(v_\text{min}) = \Theta(a - a(v_\text{min}))$, where $a(v_\text{min}) = 4 \mu_{T\chi}^2 v_\text{min}^2 /m_N^2$. We therefore refer to these functions as \emph{reduced event rates}.

The reduced event rates contain all relevant information on the experiments under consideration (such as the exposure, the energy-dependent acceptance and the energy resolution). For a given target, the functions $d_i^{(p)}(a)$ need to be calculated for $p \in \left\{ -2, -1, \dots, 15 \right\}$. Crucially, they are independent of the DM mass, the coefficient vector $\mathbf{z}$ and of the binning of $v_\text{min}$ space. Hence they only have to be calculated once for every experiment (for a given binning in recoil energy) and can then be used for arbitrary DM mass, combination of effective operators and binning of $v_\text{min}$-space.

Once the particle physics properties of DM have been specified in terms of the coefficients $\sigma_{\beta,\alpha,l}^{(\tau,\tau')}$ introduced above, it is possible to combine the reduced event rates $d_{i}^{(p)}(a)$ into two function for each bin:
\begin{align}
G_i(a) & = \frac{\kappa \, \rho}{2 \pi \,  m_\chi} \sum_{\alpha=1}^8\, \sum_{\tau=0}^1 \,\sum_{\tau'=0}^1 \,\sum_{l=-2}^5 \, \sum_{k=0}^{10} \, \sigma_{1,\alpha,l}^{(\tau,\tau')} \left(\textbf{z} \right) \, w_{\alpha,k}^{(\tau,\tau')} \, \cdot \, d_{i}^{(l+k)} (a)\,, \nonumber \\
H_i(a) & = \frac{\kappa \, \rho}{2 \pi \, m_\chi} \sum_{\alpha=1}^8\, \sum_{\tau=0}^1 \,\sum_{\tau'=0}^1 \,\sum_{l=-2}^5 \, \sum_{k=0}^{10} \, \sigma_{2,\alpha,l}^{(\tau,\tau')} \left(\textbf{z} \right) \, w_{\alpha,k}^{(\tau,\tau')} \, \cdot \, d_{i}^{(l+k)} (a)\,,
\end{align}
where the coefficients $w_{\alpha,k}^{(\tau,\tau')}$ contain the details of the nuclear response (see above).

Finally, once the binning $v_1,\ldots,v_{N_\mathrm{s}}$ of $v_\text{min}$-space has been specified, one simply needs to calculate the corresponding values $a_j \equiv a(v_j)$. The matrix $G_{ij}$ is then given by\footnote{Note that for a target consisting of different isotopes (or in fact different elements), the matrix $G_{ij}$ needs to be calculated for each isotope. The different matrices can then simply be multiplied with the mass fraction of each isotope and summed up.}
\begin{align}
G_{ij} = G_{i} \left( a_{j+1} \right) -G_{i}\left( a_j\right)
\end{align}
and the matrix $H_{ij}$ in analogy. The key point is that the second and third step require no further numerical integration and can therefore be performed very fast even for complicated non-standard interactions. Splitting the calculation of $G_{ij}$ and $H_{ij}$ into the three steps discussed above therefore allows to scan over the parameters characterising the particle physics properties of DM in a very efficient way.

\section{Calculating event rates from the velocity integral}

\subsection{The second velocity integral}
\label{ap:hcalculation}

We consider the case that the velocity integral $g(v_\text{min})$ is piecewise constant, so that it can be written as
\begin{equation}
 g(v_\text{min}) = \sum_{j=1}^{N_\mathrm{s}} l_{j} \, \Theta(v_{j+1} - v_\text{min}) \; ,
\end{equation}
where $\Theta(x)$ is the Heaviside step function and $l_j \equiv g_j - g_{j+1}$ (with $g_{N_\mathrm{s} + 1} = 0$). In order to calculate $h(v_\text{min})$ for given $v_\text{min}$, we choose $k$ in such a way that $v_{k-1} < v_\text{min} < v_k$. One then obtains:
\begin{align}
 h(v_\text{min}) & = \left[-g(v) \, v^2\right]_{v_\text{min}}^\infty + \int_{v_\text{min}}^\infty 2\,v\,g(v) \mathrm{d}v \nonumber \\
& = \sum_{j=1}^{N_\mathrm{s}} l_j \left[-\Theta(v_{j+1} - v) \, v^2\right]_{v_\text{min}}^\infty + \sum_{j=1}^{N_\mathrm{s}} l_j \int_{v_\text{min}}^\infty 2\,v\,\Theta(v_{j+1} - v) \mathrm{d}v \nonumber \\
& = \sum_{j=k-1}^{N_\mathrm{s}} l_j v_\text{min}^2 + \sum_{j=k-1}^{N_\mathrm{s}} l_j \int_{v_\text{min}}^{v_{j+1}} 2v \, \mathrm{d}v \nonumber \\
& = \sum_{j=k-1}^{N_\mathrm{s}} l_j v_\text{min}^2 + \sum_{j=k-1}^{N_\mathrm{s}} l_j \left(v_{j+1}^2 - v_\text{min}^2\right) \nonumber \\
& = \sum_{j=k-1}^{N_\mathrm{s}} l_j \, v_{j+1}^2 \; .
\end{align}
We observe that $h(v_\text{min})$ is also a piecewise-constant function, i.e.\ we can write $h(v_\text{min}) = h_j$ for $v_\text{min} \in \left[v_j,\,v_{j+1}\right]$ with
\begin{equation}
h_j = \sum_{j' = j}^{N_\mathrm{s}} l_{j'} \, v_{j'+1}^2 = \sum_{j' = j}^{N_\mathrm{s}} (g_{j'} - g_{j'+1}) \, v_{j'+1}^2 \; .
\end{equation}
The relationship between $h_j$ and $g_{j'}$ can be written as $h_j = F_{jj'} g_{j'}$ with $F_{jj} = v_{j+1}^2$, $F_{jj'} = 0$ for $j > j'$ and $F_{jj'} = v_{j'+1}^2 - v_{j'}^2$ for $j' > j$.

Using the matrix $F_{jj'}$ and the matrices $G_{ij}$ and $H_{ij}$ defined in eqs.~(\ref{eq:Gij}) and~(\ref{eq:Hij}), we can then construct the matrix $D_{ij}$ that relates the velocity integral to the predicted number of events in a given bin:
\begin{equation}
R_i = \sum D_{ij} \, g_j \quad \text{with} \quad D_{ij} = G_{ij} + \sum_{k} F_{kj} H_{ik} \; .
\end{equation}

\subsection{The impact of high-velocity bins}
\label{ap:highv}

Let us now discuss how to determine the range of $v_\text{min}$-space that should be used to calculate the matrices $G_{ij}$ and $H_{ij}$. We note that both $G_{ij}$ and $H_{ij}$ are non-zero only if the energy $E_i$ of the bin under consideration is comparable to the physical recoil energy $E_j$ corresponding to the $j$th bin in $v_\text{min}$-space. How close $E_i$ and $E_j$ have to be in order for $G_{ij}$ and $H_{ij}$ to be non-zero depends on the energy resolution $\Delta E_\mathrm{R}$, which determines the probability that a physical recoil energy $E_j$ leads to an observed energy $E_i$. The important point is that there is always a $v_\text{min}$ and a $v_\text{max}$ such that $G_{ij} \approx 0$ and $H_{ij} \approx 0$ for $v_j < v_\text{min}$ or $v_j > v_\text{max}$. This observation suggests that one can simply take $v_1 = v_\text{min}$ and $v_{N_\mathrm{s} + 1} = v_\text{max}$.

However, things become more complicated when the matrices $G_{ij}$ and $H_{ij}$ are combined into the matrix $D_{ij} = G_{ij} + \sum_{j'} H_{ij'} F_{j'j}$. To illustrate the problem, let us consider a range of $v_\text{min}$-space $[v_{1}, v_{N_\mathrm{s}+1}]$ such that $E_\mathrm{R}(v_{N_\mathrm{s}+1})$ is significantly larger than $v_\text{max}$. In this case there is a $j_\text{max}$ such that for $J > j_\text{max}$ one obtains $G_{iJ} = 0$ and $H_{iJ} = 0$ for all bins $i$. For such a $J$ it then follows that
\begin{equation}
D_{iJ} = \sum_{j'=1}^{N_\mathrm{s}} H_{ij'} F_{j'J} = \sum_{j'=1}^{j_\text{max}} H_{ij'} F_{j'J} = (v_{J+1}^2 - v_J^2) \sum_{j'=1}^{j_\text{max}} H_{ij'} \; ,
\end{equation}
because $F_{j'j} = v_{j+1}^2 - v_j^2$ independent of $j'$ for $j > j'$. Crucially we find that $D_{iJ}$ can be non-zero even for $J > j_\text{max}$.

The calculation above shows that arbitrarily high bins in $v_\text{min}$-space can influence the predictions even for the lowest bins in reconstructed energy. This observation seems paradoxical at first, but it is a direct consequence of the relation between the velocity integrals $g(v_\text{min})$ and $h(v_\text{min})$, which imply that a local variation in $g(v_\text{min})$ leads to a global variation in $h(v_\text{min})$, see also~\cite{DelNobile:2013cva}. To be more specific, if we vary $g(v_\text{min})$ only in the interval $\left[v_j, v_{j+1}\right]$ and keep it fixed everywhere else, $h(v_\text{min})$ will vary for all velocities $v_\text{min} < v_{j+1}$ and the matrix $F_{ij}$ encodes this behaviour.

At first sight, the observation that $D_{ij}$ remains non-zero for arbitrarily large $v_j$ seems to imply that an arbitrarily large number of steps is needed for a fully general treatment, which would pose a significant problem for our approach. The simplest solution to this problem would be to impose some upper limit on the allowed velocity range, for example motivated by the observed galactic escape velocity or simply by the requirement that DM ought to be non-relativistic. As we will now show, however, it is possible to solve this problem more elegantly.

The important observation is that $D_{iJ}$ takes a particularly simple form. Defining $H_i$ as the sum of all entries in the $i$th row of $H_{ij}$, i.e.\ $H_i \equiv \sum_j H_{ij}$, we obtain
\begin{equation}
D_{iJ} \equiv (v_{J+1}^2 - v_J^2) H_i \; .
\end{equation}
In other words, for fixed $J > j_\text{max}$ the column vector $D_{iJ}$ is simply proportional to the vector $H_i$.

As a result, for any vector $(g_j) = \left(g_1,\ldots,g_{N_\mathrm{s}}\right)$ in $v_\text{min}$-space we can define a new vector $(\tilde{g}_j) = \left(g_1,\ldots,g_{j_\text{max}},\tilde{g}, 0, \ldots, 0\right)$ with
\begin{equation}
 \tilde{g} = \frac{1}{(v_{j_\text{max}+2}^2 - v_{j_\text{max}+1}^2)} \sum_{J = j_\text{max} + 1}^{N_\mathrm{s}} (v_{J+1}^2 - v_J^2) g_J
\end{equation}
such that $\sum_j D_{ij} g_j = \sum_j D_{ij} \tilde{g}_j$. While $(g_j)$ is assumed to be monotonically decreasing (i.e.\ $g_{j+1} \leq g_j$), this no longer needs to be true for $(\tilde{g}_j)$, because it is possible that $\tilde{g} > g_{j_\text{max}}$. In other words, adding a large number of monotonically decreasing bins beyond $j_\text{max}$ is equivalent to adding a single bin which does not need to satisfy the monotonicity requirement. This observation implies that it is not necessary to consider arbitrarily large values of $v_\text{min}$ in order to find the best-fit velocity integral.

\providecommand{\href}[2]{#2}\begingroup\raggedright\endgroup

\end{document}